\documentclass[floatfix,prc,aps,twocolumn]{revtex4}
\usepackage{amssymb}
\usepackage{graphicx,psfrag}
\begin{document}

\newcommand{\note}[1]{\marginpar{\tiny {#1}}}	
\newcommand{\bld}[1]{\mbox{\boldmath $#1$}}	
\newcommand{\code}{$\mathtt{FREYA}$}
\newcommand{\beq}{\begin{eqnarray}}
\newcommand{\eeq}{\end{eqnarray}}
\newcommand{\SKIP}[1]{}
\newcommand{\nubar}{{\overline{\nu}}}

\title{Event-by-event evaluation of 
the prompt fission neutron spectrum from $^{239}\textrm{Pu}(n,f)$}

\author{R.~Vogt$^{1,2}$, J.~Randrup$^3$, D.~A.~Brown$^1$, 
M.~A.~Descalle$^1$, and W.~E.~Ormand$^{1,4}$}

\affiliation{
$^1$Physics Division, Lawrence Livermore National Laboratory, Livermore, CA
94551, USA\break
$^2$Physics Department, University of California, Davis, CA 95616, USA\break
$^3$Nuclear Science Division, Lawrence Berkeley National Laboratory, 
Berkeley, CA 94720, USA\break
$^4$Physics Department, Michigan State University, East Lansing, MI 48824, USA}

\date{\today}

\begin{abstract}
Earlier studies of $^{239}$Pu($n,f$) have been extended
to incident neutron energies up to 20 MeV
within the framework of the event-by-event fission model \code,
into which we have incorporated multichance fission 
and pre-equilibrium neutron emission.
The main parameters controlling prompt fission neutron evaporation 
have been identified and the prompt fission neutron spectrum 
has been analyzed by fitting those parameters 
to the average neutron multiplicity $\nubar$ from ENDF-B/VII.0, 
including the energy-energy correlations in $\nubar(E)$
obtained by fitting to the experimental $\nubar$ data 
used in the ENDF-B/VII.0 evaluation.  
We present our results, 
discuss relevant tests of this new evaluation, 
and describe possible further improvements.
\end{abstract}

\maketitle

\section{Introduction}

Nuclear fission forms a central topic in nuclear physics,
presenting many interesting issues for both experimental and theoretical 
research, and it has numerous practical applications as well,
including energy production and security.
Nevertheless, a quantitative theory of fission is not yet available.
While there has been considerable progress in the last few years,
both in liquid-drop model-type calculations \cite{MMSI-Nature409,MollerPRC79}
and in microscopic treatments \cite{BergerNPA428,GouttePRC71,DubrayPRC77},
these treatments primarily address ``cold'' fission, induced by thermal 
neutrons, and cannot yet describe ``hot fission'', induced by more energetic 
neutrons.  In order to perform new evaluations of observables important for 
applications over the full relevant energy range, it is therefore necessary 
to rely on a considerable degree of phenomenological modeling.

One of the most important quantities for applications 
is the prompt fission neutron spectrum (PFNS).  
As discussed earlier \cite{VRPY},
the experimental spectral data themselves are neither sufficiently accurate 
nor of sufficiently consistent quality to allow an improved PFNS evaluation.  
However, by combining
measured information about the nuclear fragment yields and energies with the
very precise evaluations of neutron multiplicities,
it is possible to constrain the neutron spectrum rather tightly 
without having to rely on the spectral data themselves.

Our approach employs the fission model \code\
(Fission Reaction Event Yield Algorithm)
which incorporates the relevant physics 
and contains a few key parameters that are determined 
by comparison to pertinent data through statistical analysis \cite{VRPY,RV}.
It simulates the entire fission process and produces a large sample of
complete fission events with full kinematic information on the emerging
fission products and the emitted neutrons and photons.
It thus incorporates the pre-fission emission of neutrons 
from the fissile compound nucleus
as well as the sequential neutron evaporation from the fission fragments.
\code\ provides a means of using readily-measured observables 
to improve our understanding of the fission process and
it is, therefore, a potentially powerful tool for bridging the gap 
between current microscopic models and important fission observables 
and for improving estimates of the fission characteristics 
important for applications.  

In the following, we briefly describe the employed version of \code\ 
and the fitting procedure used to obtain our extended evaluation 
of the $^{239}{\rm Pu}(n,f)$ prompt fission neutron spectrum.
We then compare our results to the ENDF-B/VII.0 \cite{ENDFb7}
evaluation of the PFNS and some benchmark criticality tests.  
Finally, we discuss the energy and model dependence of several
relevant observaables.

\section{Generation of fission events}
\label{model}

We have adapted the recently developed fission model \code\ \cite{RV}
for the present purpose of calculating the neutron spectrum 
in terms of a set of well-defined model parameters.
We describe its main physics ingredients below,
with an emphasis on the new features added for the present study,
particularly multichance fission and pre-equilibrium emission.
Being a simulation model,
\code\ follows the temporal sequence of individual fission events
from the initial agitated fissionable nucleus, 
$^{240}{\rm Pu}^*$ in the present case,
through possible pre-fission emissions
to a split into two excited fragments
and their subsequent sequential emission of neutrons and photons.
The description below is similarly organized.

\subsection{Pre-fission neutron emission}
\label{prefiss}

At low incident neutron energies, below a few MeV,
the neutron is absorbed into the target nucleus
resulting in an equilibrated compound nucleus
which may have a variety of fates.
Most frequently, in the present case, it will fission directly.
But, since the compound nucleus was formed by neutron absorption,
it is energetically possible for it to re-emit a neutron.
In that circumstance, the daughter nucleus cannot fission
and will de-excite by sequential photon emission.
\code\ generally discards such events 
because it is designed to provide fission events
(but their frequency is noted).
Neutron evaporation from a fissionable compound nucleus can be treated 
in the same manner as neutron evaporation from fission fragments,
as will be described later (Sect.\ \ref{evap}).
In principle, it is also possible that the compound nucleus will start by
radiating a photon but the likelihood for this is very small 
and is ignored.

\subsubsection{Multichance fission}
\label{nthchance}

\begin{figure}[tp]
\includegraphics[angle=270,width=\columnwidth]{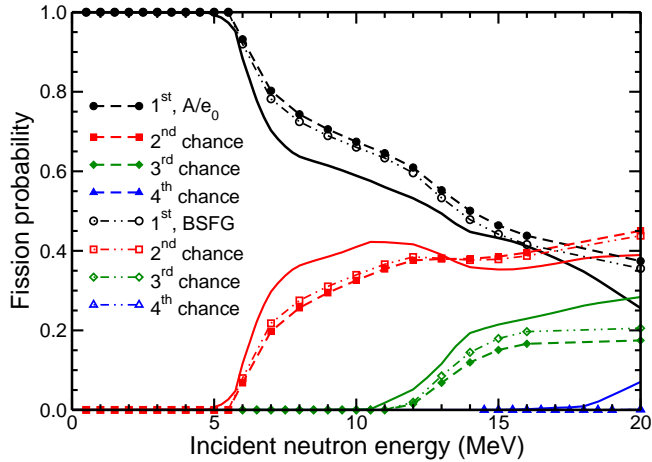}
\caption[]{(Color online)
The probability for first- (black circles), second- (red squares), third- 
(green diamonds),
and fourth- (blue triangles) chance fission as a function of 
incident neutron energy.  
The solid curves show the $\mathtt{GNASH}$ results
used in the ENDF-B/VII.0 evaluation \cite{ENDFb7},
while the dashed and dot-dot-dashed curves with closed and open symbols 
are the \code\ results discussed in the text.
}\label{f:Pnth}
\end{figure}

As the energy of the incident neutron is raised,
neutron evaporation from the produced compound nucleus competes
ever more favorably with direct (first-chance) fission.
The associated probability is given by the ratio of the fission and
evaporation widths $\Gamma_{\rm f}(E^*)$ and $\Gamma_n(E^*)$ 
for which we use the transition-state estimate \cite{SwiateckiPRC78},
\begin{equation}
{\Gamma_n(E^*) \over \Gamma_{\rm f}(E^*)}\ 
=\ 	{2g_{\rm n}\mu_n\sigma\over\pi\hbar^2}
	{\int_0^{X_n} (X_n-x)\rho_n(x)dx 
\over	 \int_0^{X_{\rm f}}		 \rho_{\rm f}(x)dx}\ ,
\end{equation}
where $g_s = 2$ is the spin degeneracy of the neutron, 
$\mu_n$ is its reduced mass,
and $\sigma=\pi R^2=\pi r_0^2A^{2/3}$.
Furthermore, $\rho_n(x)$ is the level density in 
the evaporation daughter nucleus at the excitation energy $x$,
whose maximum value is given by $X_n=Q_n=E^*-S_n$,
where $Q_n$ is the $Q$ value for neutron emission
and $S_n$ is the neutron separation energy.
Similarly, $\rho_{\rm f}(x)$ is the level density of the
transition configuration for the fissioning nucleus, {\em i.e.}\ 
when its shape is that associated with the top of the fission barrier;
the excitation $x$ is measured relative to that barrier top,
so its maximum value is $X_{\rm f}=E^*-B_{\rm f}$,
where $B_{\rm f}$ is the height of the fission barrier
(the corresponding quantity for neutron emission
is the neutron separation energy $S_n$).

Neutron evaporation is possible whenever the excitation energy of the compound
nucleus exceeds the neutron separation energy, $E_i^* > S_n$.  
(Since it costs energy to remove a neutron from the nucleus,
$S_n$ is positive.)
The excitation of the evaporation daughter nucleus 
is $E_f^* = E_i^*-S_n - E$ where $E$ is the kinetic energy 
of the relative motion between the emitted neutron and the daughter nucleus.
If this quantity exceeds the fission barrier in the daughter nucleus,
then second-chance fission is possible.
(We use the Hill-Wheeler expression for the transmission probability,
$P_f=1/[1+\exp(2\pi(B_f-E_f^*)/\hbar\omega)]$ with $\hbar\omega$=1~MeV,
so there is an exponentially small probability for sub-barrier fission.)
The procedure described above is then applied to the daughter nucleus,
thus making further pre-fission neutron emission possible.
Thus as the incident neutron energy is raised,
the emission of an ever increasing number of pre-fission neutrons
becomes possible and the associated fission events may be classified as
first-chance fission (when there are no pre-fission neutrons emitted),
second-chance fission (when one neutron is emitted prior to fission), 
and so on.

Figure~\ref{f:Pnth} shows the probabilities for $n^{\rm th}$-chance fission
obtained with \code\ for incident neutron energies up to 20~MeV, 
using two alternate assumptions about the level density parameterization.
Also shown are the $\mathtt{GNASH}$ results
used in the ENDF-B/VII.0 evaluation \cite{ENDFb7}.
The two calculations give rather similar results but,
because these probabilities are not easy to measure experimentally,
it is not possible to ascertain the accuracy of the calculations.

\subsubsection{Pre-equilibrium neutron emission}
\label{preequil}

\begin{figure}[htbp]
\includegraphics[angle=270,width=\columnwidth]{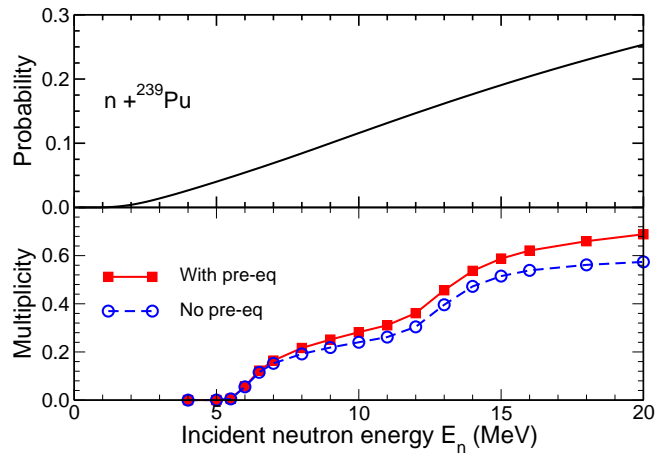}
\caption[]{(Color online)
{\em Upper panel:} The probability for pre-equilibrium neutron emission 
as a function of the incident neutron energy.
{\em Lower panel:}
The corresponding average multiplicity of neutrons emitted prior to fission
calculated without (dashed) and with (solid) the pre-equilibrium processes.
}\label{f:P-preeq}
\end{figure}


At higher incident neutron energies,
there is a growing chance that complete equilibrium is not established
before the first neutron is emitted.
Under such circumstances the calculation of statistical neutron evaporation
must be replaced by a suitable non-equilibrium treatment.
A variety of models have been developed for this process
(for example Ref.~\cite{Kawano01} which combines pre-equilibrium emission with
the Madland-Nix model \cite{MadlandNix} of the prompt fission neutron spectrum)
and we employ a practical application of the two-component exciton model
\cite{Gadioli92} (described in detail in Ref.~\cite{KoningNPA744}). 
It represents the evolution of the nuclear reaction in terms of time-dependent 
populations of ever more complex many-particle-many-hole states.
 
A given many-exciton state consists 
of $p_{\nu(\pi)}$ neutron (proton) particle excitons
and $h_{\nu(\pi)}$ neutron (proton) hole excitons.
The total number of neutron (proton) excitons in the state is 
$n_{\nu(\pi)}=p_{\nu(\pi)}+h_{\nu(\pi)}$.
The incident neutron provides the initial state
consisting of a single exciton,
namely a neutron particle excitation: $p_\nu=1$ and $p_\pi=h_\nu=h_\pi=0$.
In the course of time, the number of excitons present may change
due to hard collisions or charge exchange,
as governed by the residual two-body interaction.
We ignore the unlikely accidental processes that reduce the number of excitons,
so the state grows ever more complex.

The temporal development of the associated probability distribution
$P(p_\nu,h_\nu,p_\pi,h_\pi)$ is described by a master equation 
that accounts for the transitions between different exciton states.
The pre-equilibrium neutron emission spectrum is then given by  
\begin{eqnarray}
\frac{d\sigma_n}{dE} & = & \sigma_{\rm CN} 
\sum_{p_\pi=0}^{p_\pi^{\rm max}} 
\sum_{p_\nu=1}^{p_\nu^{\rm max}} W(p_\pi,h_\pi,p_\nu,h_\nu,E) \nonumber
\\ \mbox{} & & \, \, 
\times\tau(p_\pi,h_\pi,p_\nu,h_\nu)\, P(p_\pi,h_\pi,p_\nu,h_\nu) \, \, 
\label{preeq-spectrum}
\end{eqnarray}
where $\sigma_{\rm CN}$ is the compound nuclear cross section 
(usually obtained from an optical model calculation), 
$W$ is the rate for emitting a neutron with energy $E$ 
from the exciton state $(p_\pi,h_\pi,p_\nu,h_\nu)$, 
$\tau$ is the lifetime of this state, 
and $P(p_\pi,h_\pi,p_\nu,h_\nu)$ is the (time-averaged) 
probability for the system to survive the previous stages
and arrive at the specified exciton state.
In the two-component model, 
contributions to the survival probability 
from both particle creation and charge exchange need to be accounted for. 
The survival probability for the exciton state $(p_\pi,h_\pi,p_\nu,h_\nu)$ 
can be obtained from a recursion relation starting from the initial condition 
$P(p_\nu\!=\!1,h_\nu\!=\!0,p_\pi\!=\!0,h_\pi\!=\!0) = 1$ 
and setting $P\!=\!0$ for terms with negative exciton number.
As in Ref.~\cite{KoningNPA744},
particle emission is assumed to occur only from states with 
at least three excitons, $n_\pi +n_\nu\geq3$.
We consider excitons up to $p_\nu^{\rm max}=p_\pi^{\rm max}=6$.

The emission rate, $W(p_\pi,h_\pi,p_\nu,h_\nu,E_k)$, 
is largely governed by the particle-hole state density, 
$\omega(p_\pi,h_\pi,p_\nu,h_\nu,E^*)$. For 
a neutron ejectile of energy $E$ the rate is given by \cite{Dobes83}
\begin{eqnarray} \nonumber
&~& W(p_\pi,h_\pi,p_\nu,h_\nu,E_k)\ =\ 
\frac{g_n}{\pi^2\hbar^3} \mu_n E\,\sigma_{\rm n,inv}\\
&~&\times
\frac{\omega(p_\pi,h_\pi,p_\nu-1,h_\nu,E^*-E-S_n)}
{\omega(p_\pi,h_\pi,p_\nu,h_\nu,E^*)}\ 
\label{emission-rate}
\end{eqnarray}
where $\sigma_{k,{\rm inv}}$ is the inverse reaction cross section 
(calculated within the optical model framework) and
$E^*$ is the total excitation energy of the system.

\begin{figure}[tp]
\includegraphics[angle=270,width=\columnwidth]{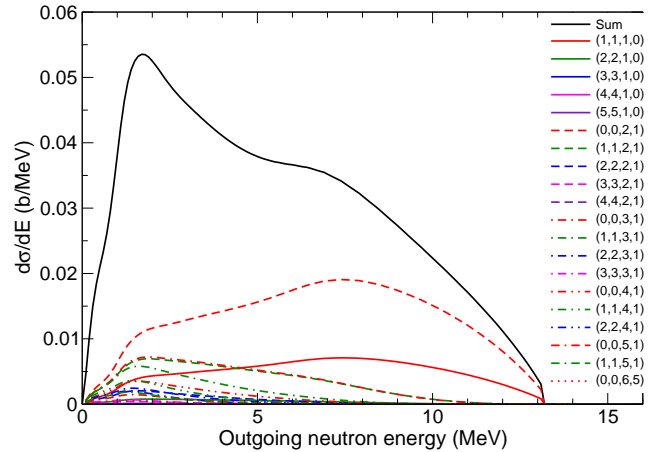}
\caption[]{(Color online)
The contributions to the pre-equilibrium neutron spectrum 
from exciton states with the indicated values of $(p_\pi,h_\pi,p_\nu,h_\nu)$,
obtained at $E_n=14$~MeV.
}\label{f:E-preeq}
\end{figure}

The calculated probability for pre-equilibrium neutron emission
is shown in the upper panel of Fig.~\ref{f:P-preeq} 
as a function of the incident neutron energy $E_n$,
while Fig.~\ref{f:E-preeq} shows the the pre-equilibrium neutron spectrum
obtained at $E_n=14$~MeV.
After being practically negligible below a few MeV, the probability for
pre-equilibrium emission grows approximately linearly to about 24\% at 20~MeV.
A careful inspection of the calculated energy spectrum shows that
neutrons emitted from states with larger exciton number approach the 
statistical emission expected from a compound nucleus,
thus ensuring our treatment has included sufficient complexity 
to exhaust the pre-equilibrium mechanism.
Because of the (desired) insensitivity to the maximum specified exciton number,
the probability shown in Fig.~\ref{f:P-preeq} is not indicative of the
importance of the pre-equilibrium processes.
Their quantitative significance is better seen by comparing
the neutron spectrum obtained with and without the pre-equilibrium treatment,
as shown in the lower panel of Fig.~\ref{f:P-preeq}.

The reaction cross sections used in Eqs.~(\ref{preeq-spectrum}) and 
(\ref{emission-rate}) define the overall magnitude of the cross sections for 
the pre-equilibrium processes. The highest accuracy results are best obtained 
from coupled-channels calculations with an appropriately-determined optical 
potential.  However, since \code\ principally deals with probabilities, the 
relative fraction of pre-equilibrium neutrons may be computed with sufficient 
accuracy employing a spherical optical potential to calculate the relevant 
cross sections. Consequently, the compund-nucleus cross sections and inverse 
cross sections were computed using the optical-model program $\mathtt{ECIS06}$ 
and the global optical model potential of Koning and 
Delaroche~\cite{KoningNPA713}. 

For each event generated, \code\ first considers the possibility of
pre-equilibrium neutron emission and, if it occurs, a neutron is emitted 
with an energy selected from the calculated pre-equilibrium spectrum
(see Fig.~\ref{f:E-preeq}).
Subsequently, the possibility of equilibrium neutron evaporation is
considered, starting either from the originally agitated compound nucleus,
$^{240}{\rm Pu}^*$, or the less excited nucleus, $^{239}{\rm Pu}^*$, 
remaining after pre-equilibrium emission has occurred.
Neutron evaporation is iterated until the excitation energy 
of a daughter nucleus is below the fission barrier
(in which case the event is abandoned and a new event is generated)
or the nucleus succeeds in fissioning.

\subsection{Mass and charge partition}
\label{split}

After the possible pre-fission processes, we are presented with a
fission-ready compound nucleus $^{A_0}Z_0$ having an excitation energy $E_0^*$.
The first task is to divide it into a heavy fragment $^{A_H}Z_H$ 
and a complementary light fragment $^{A_L}Z_L$. 
Since no quantitatively useful model is yet available for the
calculation of the fission fragment mass yields,
we have to invoke experimental data.
We follow the procedure employed in the original version of \code\
\cite{RV}.

We thus assume that the mass yields $Y(A_f)$ of the fission fragments 
exhibit three distinct modes,
each one being of Gaussian form \cite{g-fit},
\begin{equation}\label{eq:defs1s2}
Y(A_f)\ =\ S_1(A_f)+S_2(A_f)+S_L(A_f)\ .
\end{equation}
The first two terms represent asymmetric fission modes
associated with the spherical shell closure at $N=82$
and the deformed shell closure at $N=88$, respectively,
while the last term represents a broad symmetric mode, 
referred to as super-long \cite{Brosa}.
Although the symmetric mode is relatively insignificant 
at low excitations, its importance increases with the excitation energy
and ultimately dominates the mass distribution.

The asymmetric modes have a two-Gaussian form,
\begin{eqnarray}
S_i = {{N}_i\over\sqrt{2\pi}\sigma_i}\left[
 {\rm e}^{-(A_f-\bar{A}-D_i)^2/2\sigma_i^2}\!
+{\rm e}^{-(A_f-\bar{A}+D_i)^2/2\sigma_i^2}
\right]\! ,
\label{s1s2}
\end{eqnarray}
while the symmetric mode is given by a single Gaussian
\begin{eqnarray}
S_L = {N_L\over\sqrt{2\pi}\sigma_L}\,
	{\rm e}^{-(A_f-\bar{A})^2/2\sigma_L^2}\ ,
\label{slong}
\end{eqnarray}
with $\bar{A}=\mbox{$1\over2$}A_0$.
Since each event leads to two fragments, the yields are normalized so that 
$\sum_AY(A)=2$.  Thus,
\begin{eqnarray} 
2{N}_1+2{N}_2+{N}_L=2 \, \, ,
\label{nsum}
\end{eqnarray}
apart from a negligible correction 
because $A_f$ is discrete and bounded from both below and above.

Most measurements are of fission \emph{product} yields \cite{EnglandRider},
the yields after prompt neutron emission is complete.  
However, \code\ requires fission \emph{fragment} yields,
\emph{i.e.}\ the probability of a given mass partition 
before neutron evaporation has begun.  
While no such data are yet available for Pu, 
there exist more detailed data for $^{235}\textrm{U}(n,f)$
that give the fragment yields as functions of both mass and total kinetic 
energy, $Y(A_f,{\rm TKE})$ for $E_n \leq 6$ MeV \cite{HambschNPA491}.
Guided by the energy dependence of these data, 
together with other available data on the product yields 
from $^{235}$U$(n,f)$ \cite{GlendeninPRC24} 
and $^{239}$Pu$(n,f)$ \cite{GindlerPRC27}
we derive an approximate energy dependence of the fragment yields
for $^{239}$Pu$(n,f)$ up to $E_n=20$~MeV.

We now discuss how we obtained the values of the parameters used in 
Eqs.~(\ref{s1s2}) and (\ref{slong}).  
The displacements, $D_i$, away from symmetric fission in Eq.~(\ref{s1s2}) are 
anchored above the symmetry point by the spherical and deformed shell closures
and because these occur at specific neutron numbers, 
we assume that the values of $D_i$ are energy independent.  
The fitted values of the displacements for
$^{235}$U($n,f$) are $D_1 = 23.05$ and $D_2 = 16.54$.  
The values of $D_i$ should be smaller for $^{239}$Pu 
than for $^{235}$U, $D_i^{\rm U} - D_i^{\rm Pu} \approx 2$,
because the larger Pu mass is closer to the shell closure locations.  
We take $D_1 = 20.05$ and $D_2 = 14.54$ for first-chance fission 
($A_0$=240) and increase those values by $\mbox{$1\over2$}$
for each pre-fission neutron emitted.

The widths of the asymmetric modes, $\sigma_i$, 
are expanded to second order in energy:
$\sigma_i = \sigma_{i0} + \sigma_{i1} E_n
+ \sigma_{i2} E_n^2$.  We fix the energy dependence of $\sigma_i$ from
the $^{235}$U($n,f$) fragment yields as a function of mass and total kinetic 
energy of Ref.~\cite{HambschNPA491}.
To adjust our results for $^{235}$U$(n,f)$ to Pu, we assume that general
energy dependence of the parameters is the same even though the values at
$E_n = 0$ are different.  We find
\begin{eqnarray}
\sigma_1\!\! &=&\!5.6 + 0.0937 \, (E_n/{\rm MeV}) + 0.034\, (E_n/{\rm MeV})^2,\
\\
\sigma_2\!\! &=&\!2.5 + 0.11060\, (E_n/{\rm MeV}) + 0.008\, (E_n/{\rm MeV})^2.\
\end{eqnarray}
When the fissioning nucleus is the original system, $^{240}$Pu,
then $E_n$ is the value of the actual incident neutron energy.
But when the fissioning nucleus is lighter,
{\em i.e.}\ when $\nu_0$ pre-fission neutrons have been emitted,
then $E_n$ is the equivalent incident neutron energy,
{\em i.e.}\ the incident energy that would generate 
the given excitation energy $E_0^*$
when absorbed by the nucleus $^{239-\nu_0}$Pu.
The width of the super-long component, $\sigma_L$, is assumed to be constant,
independent of both the incident energy and the fissioning isotope.
We take $\sigma_L = 12$.  

The normalizations $N_i$ change only slowly with incident energy until 
symmetric fission becomes more probable, after which they decrease rapidly.
We therefore model the energy dependence of $N_i$ by a Fermi distribution,
\begin{eqnarray}
N_i = N_i^0\, (1 + \exp[(E_n - E_1)/E_2])^{-1} \, \, .
\label{normdef}
\end{eqnarray}
We assume that the midpoint and the width are the same for both
modes, $E_1 = 10.14$ MeV and $E_2 = 1.15$ MeV, so that the relative 
normalizations for the asymmetric modes have the same energy dependence.
We have not assumed that $E_1$ and $E_2$ are identical for U and Pu,
because the transition from asymmetric to more symmetric fission is not as 
smooth a function of energy in the few-MeV region for Pu as it is for U 
\cite{GlendeninPRC24,GindlerPRC27}.
We take $N_1^0 = 0.757$ and $N_2^0 = 0.242$.  With $N_1$ and $N_2$ given by
Eq.~(\ref{normdef}), $N_L$ is determined from Eq.~(\ref{nsum}).

\begin{figure}[tp]
\includegraphics[angle=270,width=\columnwidth]{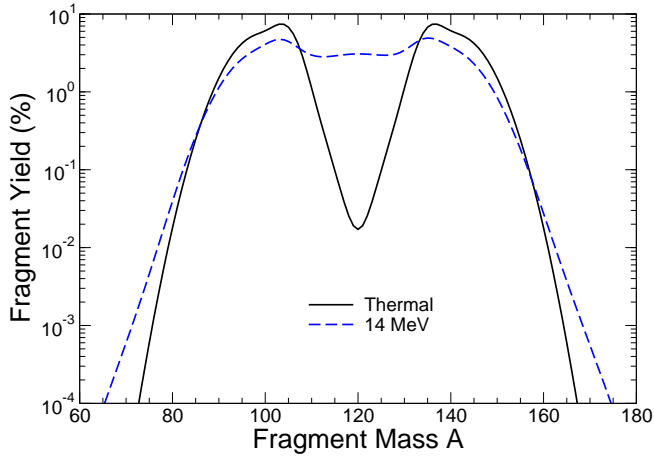}
\caption[]{(Color online)
Calculated fragment yields as  function of fragment mass for thermal 
(solid) and 14 MeV (dashed) neutrons.  The 14 MeV result also includes
contributions from multichance fission events.
}\label{yofa}
\end{figure}

The resulting fragment yields for two representative incident neutron energies
are shown in Fig.~\ref{yofa}.
The deep dip at $\mbox{$1\over2$}A_0$ visible for the thermal yields 
has substantially filled in by $E_n\! =\! 14$ MeV.  
The fragment yield at 14 MeV is a composite distribution because 
there are substantial contributions from second- and third-chance fission 
for incident neutrons of such high energy (see Fig.~\ref{f:Pnth}).  
The dashed curve in Fig.~\ref{yofa} includes these contributions 
weighted with the appropriate probabilities shown in Fig.~\ref{f:Pnth}.

The overall broadening of the yields is due in part to the larger widths
of the $S_1$ and $S_2$ modes at higher energies and in part to the increased
contribution of the $S_L$ component.  We note that while $\sigma_L$ does not
change, the larger $N_L$ enhances the importance of this component.

Once the Gaussian fit has been performed,
it is straightforward to make a statistical selection of a fragment
mass number $A_f$.  The mass number of the partner fragment is then readily
determined since $A_L+A_H=A_0-\nu_0$.

The fragment charge, $Z_f$, is selected subsequently.
For this we follow Ref.~\cite{ReisdorfNPA177} and employ a Gaussian form,
\begin{equation}
P_{A_f}(Z_f)\ \propto\ {\rm e}^{-(Z_f-\bar{Z}_f(A_f))^2/2\sigma_Z^2}\ ,
\end{equation}
with the condition that $|Z_f-\bar{Z}_f(A_f)|\leq5\sigma_Z$.
The centroid is determined by requiring that the fragments have, on average,
the same charge-to-mass ratio as the fissioning nucleus,
$\bar{Z}_f(A_f)=A_fZ_0/A_0$.  The dispersion is 
the measured value, $\sigma_Z=0.5$ \cite{ReisdorfNPA177}.
The charge of the complementary fragment then follows using $Z_L+Z_H=Z_0$.

\subsection{Fragment energies}

Once the partition of the total mass and charge among the two fragments
has been selected, 
the $Q$ value associated with that particular fission channel follows 
as the  difference between the total mass of the fissioning nucleus
and the ground-state masses of the two fragments,
\begin{equation}
Q_{LH}\ =\ M(^{240-\nu_0}{\rm Pu}^*) - M_L - M_H\ .
\end{equation}
\code\ takes the required nuclear ground-state masses
from the compilation by Audi {\em et al.}~\cite{Audi},
supplemented by the calculated masses of M{\"o}ller {\em et al.}~\cite{MNMS}
when no data are available.
The $Q_{LH}$ value for the selected fission channel is then divided up between
the total kinetic energy (TKE) and the total excitation energy (TXE)
of the two fragments.
The specific procedure employed is described below.

Through energy conservation,
the total fragment kinetic energy TKE is intimately related to the
resulting combined multiplicity of evaporated neutrons, $\nu_L+\nu_H$,
which needs to be obtained very accurately.

\begin{figure}[tp]
\includegraphics[angle=270,width=\columnwidth]{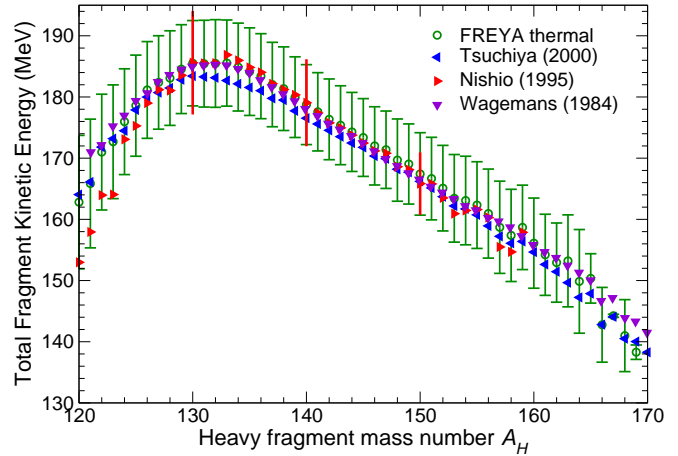}
  \caption[]{(Color online)
The measured average TKE as a function of the mass number of
the heavy fragment \protect\cite{WagemansPu,NishioPu,TsuchiyaPu} 
compared to \code\ calculations at thermal energies.  The \code\ result 
is shown 
with the calculated variance around each $A$.}
\label{f:TKEexp}
\end{figure}

Figure~\ref{f:TKEexp} shows the measured average TKE value
as a function of the mass number of the heavy fragment, $A_H$.
Near symmetry, the plutonium fission fragments are mid-shell nuclei
subject to strong deformations.  Thus the scission configuration will contain
significant deformation energy and a correspondingly low TKE.  
At $A_H = 132$, the heavy fragment is close to the doubly-magic closed shell 
having $Z_H = 50$ and $N_H = 82$ 
and is therefore resistant to distortions away from sphericity.
Consequently, the scission configuration is fairly compact,
causing the TKE to exhibit a maximum
even though the complementary light fragment is far from a closed shell 
and hence significantly deformed.

The TKE values shown in Fig.~\ref{f:TKEexp} were obtained in experiments 
using thermal neutrons \cite{WagemansPu,NishioPu,TsuchiyaPu}. 
Unfortunately, there are no such data for higher incident energy.
We therefore assume the energy-dependent average TKE values take the form
\begin{eqnarray}
\overline{\rm TKE}(A_H,E_n)\ =\ 
	\overline{\rm TKE}_{\rm data}(A_H) + d{\rm TKE}(E_n)\ .
\label{dtkevalue}
\end{eqnarray}
The first term is extracted from the data shown in Fig.~\ref{f:TKEexp},
while the second term is a parameter adjusted to ensure reproduction
of the measured energy-dependent average neutron multiplicity,
$\overline{\nu}(E_n)$.
In each particular event, the actual TKE value is then obtained
by adding a thermal fluctuation to the above average, as explained later.

Figure \ref{f:TKEexp} includes the average TKE values
calculated with \code\ at thermal energies,
together with the associated dispersions 
(these bars are {\em not} sampling errors
but indicate the actual width of the TKE distribution for each $A_H$).

\begin{figure}[tp]
\includegraphics[angle=270,width=\columnwidth]{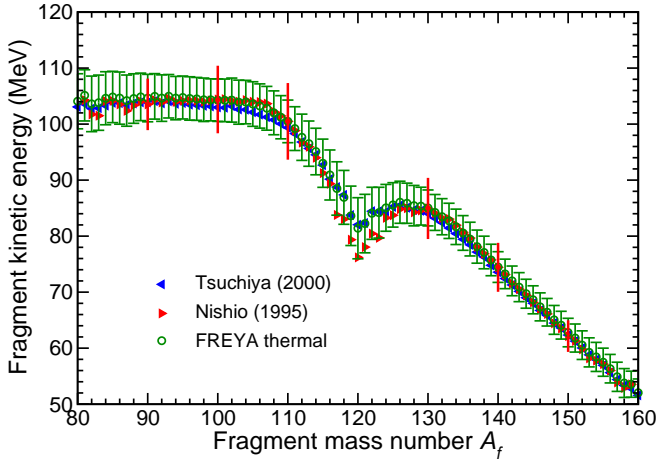}
  \caption[]{(Color online)
The average fragment kinetic energy
as a function of fragment mass from Refs.~\protect\cite{NishioPu,TsuchiyaPu}
compared to \code\ 
calculations at thermal energies.  The \code\ result is shown with the 
calculated variance around each $A$.
}\label{f:KE}
\end{figure}

Figure \ref{f:KE} shows the single-fragment kinetic energy
obtained with \code\ for incident thermal neutrons.
Although \code\ is not tuned to match the single fragment-kinetic energies, 
it does reproduce these data quite well.
Due to momentum conservation, the light fragment carries away 
significantly more kinetic energy than the heavy fragment.
Furthermore, the kinetic energy of the fragment is nearly constant 
for $A_f< 106$, but after the dip near symmetry
there is an approximately linear decrease in the fragment kinetic energy.
The figure also shows the calculated width in the fragment energy distribution,
together with a few typical experimental widths provided by 
Ref.~\cite{NishioPu}.

\begin{figure}[tp]
\includegraphics[angle=270,width=\columnwidth]{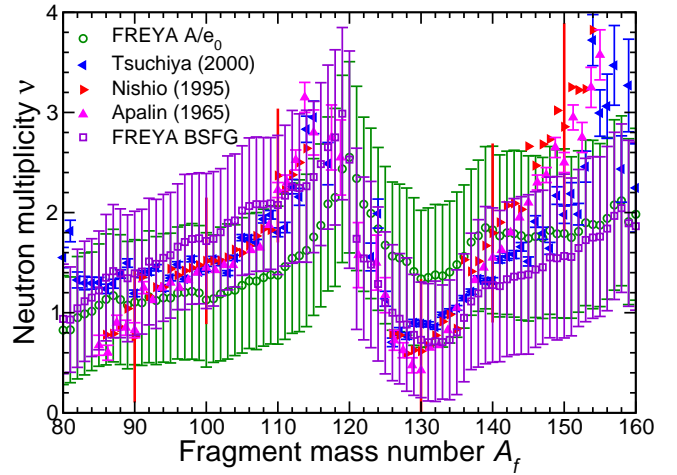}
  \caption[]{(Color online)
The measured average neutron multiplicity as a function of the fragment mass 
\protect\cite{NishioPu,TsuchiyaPu,ApalinPu} 
together with thermal \code\ results for $a = A/e_0$ (circles) and from
Eq.~(\protect\ref{aleveldef}),
with both the average and the dispersion indicated for each $A_{\rm f}$.
}\label{nuvsa_thermal}
\end{figure}

Of particular interest is the dependence of the average neutron multiplicity
on the fragment mass number $A_f$,
shown in Fig.\ \ref{nuvsa_thermal}.
It is seen that the \code\ calculations provide 
a rather good representation of the `sawtooth' behavior 
of $\bar{\nu}(A_f)$, as shown in Fig.~\ref{nuvsa_thermal},
even though \code\ is also not tuned to these data.  Although the agreement
is good, the observed behavior is not perfectly reproduced.
When $A_f>150$, a region where the fragment yield is decreasing sharply,
the data and the calculations appear to diverge.  We note that the 
uncertainties on the data in this region, where reported, are rather large.

Once the average total fragment kinetic energy has been obtained, 
the average combined excitation energy in the two fragments follows
by energy conservation,
\begin{equation}
\overline{\rm TXE}\ =\ \overline{E}_L^*+\overline{E}_H^*\
\doteq\	Q_{LH} - \overline{\rm TKE}\ .
\end{equation}
The first relation indicates that the total excitation energy
is partitioned between the two fragments.
As is common, we assume that the fragment level densities are of the form
$\rho_i(E_i^*)\sim\exp(2\sqrt{{a}_iU_i})$, 
where $U_i$ is the effective statistical energy in the fragment.

We have studied two forms of the level-density parameter, ${a}_i$.
One is the simple form $a_i=A_i/e_0$ used in the Madland-Nix model;
it has no energy dependence and may be appropriate for those
(far-from-stability) fragments for which little information is available, 
even for the ground states.  
As an alternative, we have also used a parameterization based on 
the back-shifted Fermi gas (BSFG) model \cite{Kawano},
\begin{eqnarray}
a_i({E}_i^*) = {A_i\over e_0}
\bigg[ 1 + \frac{\delta W_i}{U_i} \left(1 - e^{-\gamma U_i}\right) \bigg]\ ,
\label{aleveldef}
\end{eqnarray}
where $U_i={E}_i^*-\Delta_i$ and $\gamma=0.05$, also used in 
Ref.~\cite{LemairePRC72}.  The pairing energy of the fragment, $\Delta_i$,
and its shell correction, $\delta W_i$,
are tabulated in Ref.~\cite{Kawano} 
based on the mass formula of  Koura {\em et al.}~\cite{Koura}.
If $\delta W_i$ is negligible or if $U$ is large
then the renormalization of $A_i$ is immaterial
and the BSFG level-density parameter reverts to the simple form,
${a}_i\approx A_i/e_0$. [Because the back shift causes 
$U_i$ to become negative when $E_i^*$ is smaller than $\Delta_i$,
we replace $U_i(E_i^*)$ by a quadratic spline for $E_i^*\leq2\Delta_i$
while retaining the expressions $T_i=\sqrt{U_i/a_i}$ for the temperature and
$\sigma_{E_i}^2=2U_iT_i$ for the variance of the energy distribution
to ensure a physically reasonable behavior.]
In both scenarios, we take $e_0$ as an adjustable model parameter.

If the two fragments are in mutual thermal equilibrium, $T_L\!=\!T_H$,
the total excitation energy will, on average,
be partitioned in proportion to the respective heat capacities
which in turn are proportional to the level-density parameters,
{\em i.e.}\ $\overline{E}^*_i\sim{a}_i$.
\code\ therefore first assigns tentative average excitations
based on such an equipartition,
\begin{equation}
\acute{E}_i^*\ =\ 
{{a}_i(\tilde{E}_i^*) \over 
{a}_L(\tilde{E}_L^*)+{a}_H(\tilde{E}_H^*)}\,\overline{\rm TXE}\ ,
\end{equation}
where $\tilde{E}_i^*=(A_i/A_0)\overline{\rm TXE}$.
Subsequently, because the observed neutron multiplicities suggest
that the light fragments tends to be disproportionately excited,
the average values are adjusted in favor of the light fragment,
\begin{eqnarray}
\overline{E}^*_L = x \acute{E}^*_L\ ,\ 
\overline{E}^*_H = \overline{\rm TKE}-\overline{E}^*_L\ ,
\label{eeshift}
\end{eqnarray}
where $x$ is an adjustable model parameter expected be larger than unity,
as suggested by measurements of $^{235}$U(n,f) \cite{NishioU}
and $^{252}$Cf(sf) \cite{VorobyevCf}.

After the mean excitation energies have been assigned,
\code\ considers the effect of thermal fluctuations.
The fragment temperature $T_i$ is obtained from 
$\overline{U}_i\equiv U_i(\bar{E}_i^*)=a_iT_i^2$
and the associated variance in the excitation $E_i^*$ 
is taken as $\sigma_i^2=2\overline{U}_i^*T_i$,
where $U(E^*)=E^*$ in the simple (unshifted) scenario.

Therefore, for each of the two fragments,
we sample a thermal energy fluctuation $\delta E_i^*$ 
from a normal distribution of variance $\sigma_i^2$
and modify the fragment excitations accordingly, arriving at
\beq
E_i^*\ =\ \overline{E}_i^*+\delta E_i^*\ ,\ i=L, H .
\eeq
Due to energy conservation, there is a compensating opposite fluctuation in
the total kinetic energy, so that
\begin{eqnarray}\label{tkefinal}
{\rm TKE}\ =\ \overline{\rm TKE} - \delta E^*_L - \delta E^*_H\ .
\end{eqnarray}

With both the excitations and the kinetic energies of the two fragments
fully determined, it is an elementary matter to calculate the magnitude
of their momenta with which they emerge after having been fully accelerated 
by their mutual Coulomb repulsion \cite{RV}.
The fission direction is assumed to be isotropic 
({\em i.e.}\ directionally random) in the frame of the fissioning nucleus
and the resulting fragment velocities are finally Lorentz boosted 
into the rest frame of the original $^{240}{\rm Pu}^*$  system.

\subsection{Fragment de-excitation}

Usually both fully accelerated fission fragments are excited sufficiently 
to permit the emission of one or more neutrons.
We simulate the evaporation chain in a manner conceptually similar
to the method of Lemaire {\em et al.}\ \cite{LemairePRC72} 
for $^{252}{\rm Cf}({\rm sf})$ and $^{235}{\rm U}(n,f)$.
After neutron emission is no longer energetically possible,
\code\ simulates the sequential emission of photons by a similar method
\cite{RV}, see also Ref.~\cite{LemairePRC73}.

\subsubsection{Neutron evaporation}
\label{evap}

Neutron emission is treated 
by iterating a simple neutron evaporation procedure
for each of the two fragments separately.
At each step in the evaporation chain,
the excited mother nucleus $^{A_i}Z_i$
has a total mass equal to its ground-state mass
plus its excitation energy, $M_i^* = M_i^{\rm gs}+ E_i^*$.
The $Q$-value for neutron emission from the fragment
is then $Q_n=M_i^* -M_f - m_n$,
where $M_f$ is the ground-state mass of the daughter nucleus
and $m_n$ is the mass of the neutron.
(For neutron emission we have $A_f=A_i-1$ and $Z_f=Z_i$)
The $Q$-value is equal to the maximum possible excitation energy
of the daughter nucleus,
achieved if the final relative kinetic energy vanishes.
The temperature in the daughter fragment is then maximized at $T_f^{\rm max}$.
Thus, once $Q_n$ is known, the kinetic energy 
of the evaporated neutron may be sampled.
\code\ assumes that the angular distribution is isotropic 
in the rest frame of the mother nucleus 
and uses a standard spectral shape \cite{Weisskopf},
\beq
f_n(E)\ \equiv\ {1\over N_n}{dN_n\over dE}\
\sim\ E\,{\rm e}^{-E/T_f^{\rm max}}\ ,
\eeq
which can be sampled efficiently \cite{RV}.

Although relativistic effects are very small for neutron evaporation, they
are taken into account to ensure exact conservation of energy and momentum,
which is convenient for code verification purposes.
We therefore take the sampled energy $E$ to represent the
{\em total} kinetic energy in the rest frame of the mother nucleus,
{\em i.e.}\ it is the kinetic energy of the emitted neutron
{\em plus} the recoil energy of the excited residual daughter nucleus.
The daughter excitation is then given by
\beq
E_d^*\ =\ Q_n-E \,\, .
\eeq
and its total mass is thus $M_d^* = M_d^{\rm gs} + E_d^*$.  
The magnitude of the momenta of the excited daughter and the emitted neutron
can then be determined  \cite{RV}.
Sampling the direction of their relative motion isotropically,
we thus obtain the two final momenta
which are subsequently boosted into the overall reference frame
by the appropriate Lorentz transformation.

This procedure is repeated until 
no further neutron emission is energetically possible, 
{\em i.e.}\ when $E_d^*<S_n$,
where $S_n$ is the neutron separation energy 
in the prospective daughter nucleus,
$S_n=M(^{A_d}Z_d)-M(^{A_d-1}Z_d)-m_n$.

\subsubsection{Photon radiation}
\label{photons}

After the neutron evaporation has ceased,
the excited product nucleus may de-excite by sequential photon emission.
\code\ treats this process in a manner analogous to neutron evaporation,
{\em i.e.}\ as the statistical emission of massless particles.
While this simple treatment is expected to be fairly reasonable
at the early stage where the level density can be regarded as continuous,
it would obviously be inadequate for late stages
that involve transitions between specific levels.

There are two important technical differences relative to the treatment of 
neutron emission.
There is no separation energy for photons and, since they are massless,
there is no natural end to the photon emission chain.
We therefore introduce an infrared cutoff of 200~keV.
Whereas the neutrons may be treated by nonrelativistic kinematics,
the photons are ultrarelativistic.
As a consequence, their phase space has an extra energy factor,
\beq
f_\gamma(E)\ \equiv\ {1\over N_\gamma}{dN_\gamma\over dE}\
\sim\ E^2\,{\rm e}^{-E/T_f^{\rm max}}\ .
\eeq

The photons are assumed to be emitted isotropically
and their energy can be sampled very quickly
from the above photon energy spectrum \cite{RV}.
The procedure is repeated until the available energy is below the specified
cutoff, yielding a number of kinematically fully-characterized photons
for each of the product nuclei.

\section{Results}
\label{results}

We now proceed to discuss our analysis of the prompt fission neutron spectrum 
(PFNS). We first describe the computational approach
and then explain how the model parameters are determined.
The resulting prompt neutron spectral evaluations are then discussed in detail.
Finally, we present some additional observables of particular relevance.

\subsection{Computational approach}
\label{method}

Here we briefly describe the statistical method used
for determining model parameters and reaction observables.
Our analysis uses the Monte Carlo approach to Bayesian inference outlined 
in many books on general inverse problem theory, {\em e.g.} 
Ref~\cite{Tarantola}.

We have introduced several model parameters:
$e_0$, $x$, and $d{\rm TKE}$,
which in principle may be adjusted 
for each incident neutron energy $E_n$.
However, because independent fits to the experimental data
tend to yield values of $e_0$ and $x$ that are nearly independent
of $E_n$ \cite{VRPY}, 
we shall assume that these two parameters are energy independent.
This simplification will facilitate the optimization procedure.
Thus a given model realization is characterized by the two values $e_0$ and $x$
together with the function $d{\rm TKE}(E_n)$ which,
for practical purposes, 
will be defined by its values at certain selected energies, 
$\{d{\rm TKE}_\ell\}$.
For formal convenience, we denote the set of model parameter values
as $\bld{m}= \{m_k\}$.    

When \code\ is used with any particular value set $\bld{m}$,
it yields a sample of fission events from which we can extract observables,
$\bld{d}(\bld{m})$, that can be directly compared 
to the corresponding experimental values, $\bld{d}_{\rm exp}$.
For example, we may extract the energy-dependent mean neutron multiplicity, 
$\nubar(E_n)$, and compare it with the values given in 
the ENDF/B-VII.0 evaluation \cite{ENDFb7}.

We assume that the experiment provides not just the values 
but also the entire associated covariance matrix $\bld{\Sigma}_{\rm exp}$.
(The diagonal elements of $\bld{\Sigma}_{\rm exp}$
are the variances on the individual observables.)
The degree to which the particular model realization
defined by the parameter values $\bld{m}$ 
describes the measured data $\bld{d}_{\rm exp}$
is then expressed by
\begin{equation}
P(\bld{d}_{\rm exp}|\bld{m} ) 
\sim	\exp{\left(-\mbox{$1\over2$}\chi^2(\bld{m})\right)}\ .
\end{equation}
where $\chi^2(\bld{m})$ is the generalized least-squares deviation 
between the model $\bld{m}$ and experiment, 
\begin{equation}
\chi^2 = (\bld{d}_{\rm exp} -\bld{d}(\bld{m})) 
\cdot	(\bld{\Sigma}_{\rm exp})^{-1} 
\cdot	(\bld{d}_{\rm exp} -\bld{d}(\bld{m}))^T\ .
\end{equation}
Employing merely the diagonal part of $\bld{\Sigma}_{\rm exp}$,
{\em i.e.}\ the uncertainties alone,
ensures that well-measured observables carry more weight 
than poorly measured ones.  
This approach was used in the previous PFNS evaluation \cite{VRPY},
which was restricted to lower energy ($E_n < 5.5$~MeV).
Here we now employ the full covariance matrix, 
thereby ensuring that correlations between measured observables 
are also taken into account.
As we shall see, these correlations do impact the results.

Using the above framework, we may now compute the weighted averages 
of arbitrary observables $\bld{O} = \{{\cal O}_i\}$.
We assume that the physically reasonable values of the model parameters 
$\bld{m}$ are uniformly distributed within a hypercube in parameter space.
This defines the {\em a-priori} model probability distribution $P(\bld{m})$.
The {\em best estimate} of the observable ${\cal O}_i$ is then given by
\begin{equation}
	\prec{\cal O}_i\succ\ = 
\int d\bld{m}\,P(\bld{m})\,P(\bld{d}_{\rm exp}|\bld{m})\,{\cal O}_i(\bld{m})\ .
\end{equation}
The best estimate for the covariance between two such observables
can be obtained similarly,
\begin{eqnarray}
O_{ij}\ & \equiv\ &	\prec{\cal O}_i{\cal O}_j\succ
		-	\prec{\cal O}_i\succ \, \prec{\cal O}_j\succ\ 
\nonumber \\
        & = & \prec {\cal O}_i - \prec{\cal O}_i \succ \succ \prec {\cal O}_j
- \prec{\cal O}_j \succ \succ\ \, \, .
\label{eq:defofcov}
\end{eqnarray}
In particular, we may compute the best estimate of $\nubar$,
\begin{equation}
	\prec \overline\nu\succ\ =\
\int d\bld{m}\,P(\bld{m})\,P(\bld{d}_{\rm exp}|\bld{m})\,\nubar(\bld{m})\ ,
\end{equation}
and the prompt neutron spectrum,
as well as the covariances between those quantities.

In practice, we average over parameter space employing a Monte Carlo approach, 
thereby reducing the integral over all possible parameter values $\bld{m}$
to a sum over $N$ sampled model realizations, $\{\bld{m}^{(n)}\}$,
\begin{equation}
\prec{\cal O}_i\succ\, \approx \frac{1}{N}\! \sum_{n=1}^N  
	P(\bld{m}^{(n)})\,
	P(\bld{d}_{\rm exp}|\bld{m}^{(n)})\,{\cal O}_i(\bld{m}^{(n)}) \, \, .
\end{equation}
The joint probability 
$w_n\equiv P(\bld{m}^{(n)}) P(\bld{d}_{\rm exp}|\bld{m}^{(n)})$ may be viewed 
as the likelihood that the particular model realization $\bld{m}^{(n)}$
is ``correct.''  Since it depends exponentially on $\chi_n^2$,
the likelihood tends to be strongly peaked around the favored set.
It is important that the parameter sample be sufficiently dense
in the peak region to ensure that many sets have non-negligible weights.
We use Latin Hypercube sampling (LHS) \cite{McKay,Iman},
which samples a function of $K$ variables
with the range of each variable divided into $M$ equally-spaced intervals.  
Each combination of $M$ and $K$ is sampled at most once,
with a maximum number of combinations being $(M!)^{K-1}$.  
The LHS method generates samples that better reflect 
the distribution than a purely random sampling would.
Consequently, relative to a simple Monte Carlo sampling,
the employed sampling method requires fewer realizations
to determine the optimal parameter set.
We used 5000 realizations of the parameter space to
obtain the optimal parameter values.

Even though both $\nubar$ and the neutron spectrum are important observables,
the fact that the uncertainties on the evaluated $\nubar$ are so relatively
small drive the results.  
Indeed, we use the evaluated $\nubar$ to constrain our
new evaluation of the PFNS, as in Ref.~\cite{VRPY}.  Thus, in our 
treatment, the spectra is an outcome 
rather than a comparative observable. 
We use the evaluated $\nubar$ in the ENDF/B-VII.0 database \cite{ENDFb7}
with the covariance resulting from the least-squares fit to the available
$^{239}$Pu($n,f$) data described in Ref.~\cite{Pueval}.  
The energy-dependent neutron multiplicity, $\nubar(E_n)$
is represented as a locally linear fit to the experimental data.  
Since the nodes in this fit do not align with fitted data the fit 
introduces energy correlations that are encoded into the covariance matrix.

For each model realization,
\code\ is used to generate a large sample of fission events
(typically one million events for each parameter set) 
for each of the selected incident neutron energies	
and the resulting average multiplicity $\nubar(E_n)$
is extracted from the generated event sample.

\subsection{Fit results}
\label{fresults}

Given the tendency of $e_0$ to be larger than 8~MeV, 
we let $e_0$ vary between 8 and 12~MeV.  
Also, recalling our previous results \cite{VRPY}
and the experimental indications that the light fragment is hotter
than the heavy fragment, we have assumed {\em $1\leq x\leq 1.4$}.  
The resulting optimal values of these parameters are 
$e_0=10.0724 \pm 0.5571$~MeV and $x=1.2339 \pm 0.0496$ 
for the BSFG level density and
$e_0=9.6068 \pm 0.8930$~MeV and $x=1.0164 \pm 0.1661$ for $a = A/e_0$.
These values of $e_0$ are consistent with the calculation of $a$ in
Eq.~(\ref{aleveldef}) which does not include collective effects.  If collective
behavior is included, then we expect $e_0 \sim 13$~MeV based on the RIPL-3
systematics \cite{RIPL3}.  Previously, we obtained $e_0 \sim 8$~MeV and $x \sim
1.1$ with a slightly different TKE prescription and using only the diagonal 
elements of the $\nubar$ covariance matrix \cite{VRPY}.

\begin{figure}[tp]
\includegraphics[angle=270,width=\columnwidth]{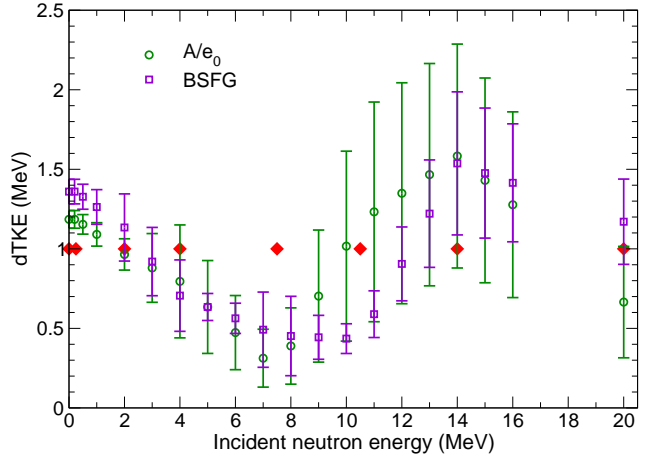}
  \caption[]{(Color online)
The fitted values of $d$TKE as a function of incident neutron energy 
for $a = A/e_0$ (circles) and the BSFG (squares).  The locations of 
the node points are indicated by diamonds at $d{\rm TKE}=1$.}
\label{dTKEfits}
\end{figure}

We have fit $d$TKE at eight values of incident neutron energy,
$E_n = 10^{-11}$, 0.25, 2, 4, 7.5, 10.5, 14 and 20 MeV,
to keep the parameter space manageable.  These points are chosen to reflect
the physics of the fission process: the region between 0.25 and 2 MeV is where
$\overline \nu(E_n)$ changes slope while 
the second and third-chance fission thresholds occur in the regions 
$4 < E_n < 7.5$ MeV and $10.5 < E_n < 14$ MeV, respectively.
The full 20-point grid of the \code\ evaluation is then covered by means of
a linear interpolation between these node points.
The resulting values of $d$TKE for the two level density prescriptions 
are shown in Fig.~\ref{dTKEfits}.  The locations of the node points are 
indicated by diamonds at $d{\rm TKE}=1$.
The error bars on $d$TKE at the node points are the standard deviations 
obtained from the averaging over the range of parameter values
while the error bars on $d$TKE between two node points 
are the interpolated dispersions between those two points.  
The standard deviations of $d$TKE are larger for the case 
where $ a = A/e_0$, possibly because the
value of $x$ is close to the edge of the fit range.

Because $d$TKE represents the shift in the total fragment kinetic energy
from the value obtained for incident thermal neutron energies,
$d$TKE should depend on the incident neutron energy, as suggested in 
Eq.~(\ref{dtkevalue}).  The value of $d$TKE is positive, indicating that using
the thermal average value of TKE leads to too many neutrons.  
The positive $d$TKE is then
required to reduce the excitation energy sufficiently to give a good fit
to $\nubar$.  
For example, reducing $d$TKE at $E_n = 0.5$~MeV from 1.4~MeV to zero
while retaining the same values of $e_0$ and $x$,
reduces $\overline{\rm TKE}$ by about 1\% and increases $\nubar$ by $\sim 6$\%.
The change in $d$TKE with energy is not particularly systematic in either
scenario.  

Above 14 MeV, the ENDF/B-VII.0 $\nubar$ evaluation is not based on data but 
on a linear extrapolation of measurements taken for higher incident energies.
Thus, $\overline \nu$ is not well constrained 
near the high end of the energy range.

\begin{figure}[tp]
\includegraphics[angle=270,width=\columnwidth]{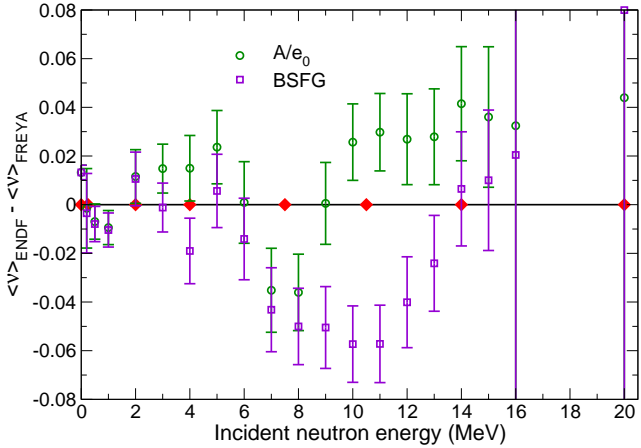}
  \caption[]{(Color online)
The difference between the ENDF-B/VII.0 evaluation 
and our fits to $\overline \nu$ using \code\ with $a = A/e_0$ (circles)
and the BSFG result (squares).  The locations of 
the node points are indicated by diamonds at $\langle \nu \rangle_{\rm ENDF}
- \langle \nu \rangle_{\mathtt{FREYA}}=0$.
}\label{nubarresidual}
\end{figure}

A direct comparison of our fitted values of $\overline \nu$ with those 
in the ENDF-B/VII.0 evaluation is less revealing than than fit residuals, 
the difference $\overline \nu_{\rm ENDF} - \overline \nu_{\mathtt{FREYA}}$.  
The residual values are shown in Fig.~\ref{nubarresidual}.  
The standard deviation on each point reflects only the uncertainty 
on the ENDF-B/VII.0 evaluation and not the uncertainty on the fitted $\nubar$.
The uncertainties on $\overline \nu$ arising from the fitting procedure 
are typically smaller than those on the evaluated values.  
The largest residuals occur for $E_n > 2$ MeV.  
The large uncertainties on the residuals at 16 and 20 MeV arise 
from the lack of experimental data at these values of $E_n$.  

We note that the ENDF-B/VII.0 $\nubar$ evaluation lies more than one standard 
deviation above the evaluated $\nubar$ 
data extracted in the ENDF-B/VII.0 covariance analysis 
in the region $0.1 < E_n < 1$ MeV \cite{Pueval}.  
Our results with both calculations of the level-density parameter 
agree rather well with the evaluated data in this region.
Here, where $\overline \nu$ is smallest, 
the relative difference is less than 0.5\%.  
At higher energies the relative uncertainty increases because, 
while $\overline \nu$ increases with incident energy, 
the residual also increases.  The residual difference is largest for the BSFG
at $6 <E_n<14$~MeV.  We note that, assuming each energy point is independent of
all others, {\em e.g.}\ the errors are uncorrelated and the covariance matrix
is diagonal, then the fit residuals can be very small, as in our previous
work \cite{VRPY}.

\begin{figure}[tp]
\begin{center}
\includegraphics[angle=270,width=\columnwidth,clip]{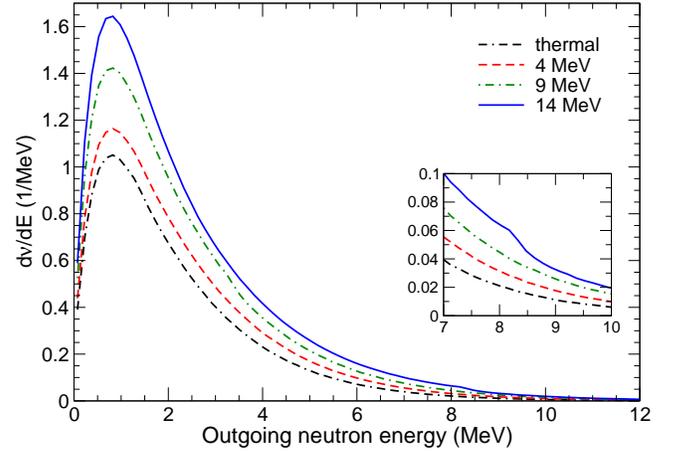}
\end{center}
\caption[]{(Color online)
The prompt neutron spectra at selected energies resulting from our BSFG fits.
}\label{nspectraBSFG}
\end{figure}

Examples of the resulting prompt fission neutron spectrum for the BSFG
level density are shown in 
Fig.~\ref{nspectraBSFG}.  We present $d\nu/dE$ 
for four representative energies:
$E_n = 0.5$ MeV (thermal neutrons), 
4~MeV (below the second-chance fission threshold),
9~MeV (below the threshold for third-chance fission),
and 14~MeV (relevant for certain experimental tests).  
We note that the integral of $d\nu/dE$ over outgoing
neutron energies gives the average neutron multiplicity, $\nubar$, 
obtained from the 
fitting procedure.  In addition to the increase in the peak of the spectrum,
we note that the average outgoing neutron energy 
appears to increase with incident energy.

\begin{figure}[tp]
\begin{center}
\includegraphics[angle=270,width=\columnwidth,clip]{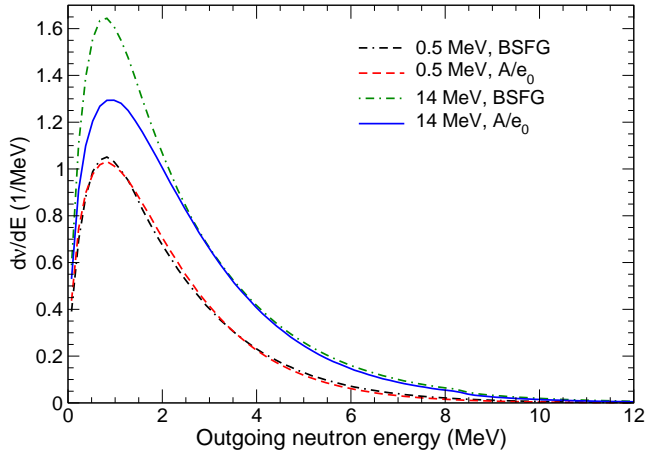}
\end{center}
\caption[]{(Color online)
Comparison of the spectra resulting from the BSFG and $ a = A/e_0$ level 
density parameterizations at $E_n = 0.5$ and 14 MeV.
}\label{nspectra_comp}
\end{figure}

Figure~\ref{nspectra_comp} compares results for the BSFG level
density prescription and $a = A/e_0$ at $E_n = 0.5$ and 14 MeV.  
While the peaks appear at the
same point at 0.5~MeV, the spectra with $a = A/e_0$ is broader around the peak
than the BSFG case because the back shift has the effect of narrowing the
spectrum.  This effect is particularly apparent at 14 MeV where the BSFG peak
is significantly higher than that for $a = A/e_0$ even though the residual
differences between the evaluation and the fitted $\nubar$ values in 
Fig.~\ref{nubarresidual} are negligible on this scale.

\begin{figure}[tp]
\includegraphics[angle=270,width=\columnwidth]{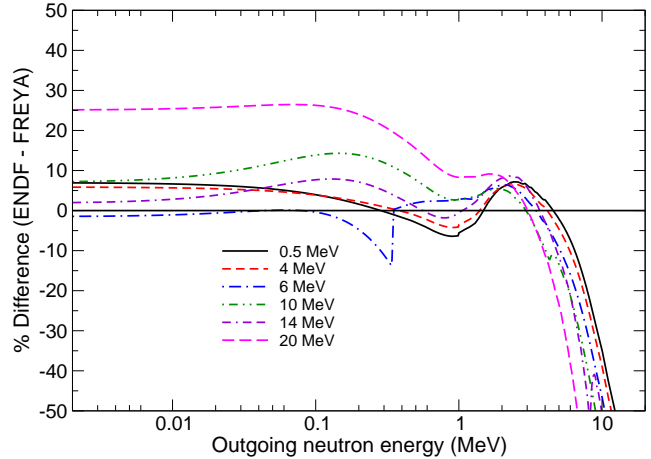}
  \caption[]{(Color online)
The percent difference between the ENDF-B/VII.0 evaluation of the PFNS and our
BSFG result at several representative incident neutron energies.  Results are
shown for 0.5 (solid black), 4 (short dashed red), 6 (dot-dashed blue), 
10 (dot-dot-dashed green), 14 (dot-dash-dashed violet) and 20 (long dashed 
magenta) MeV.}
\label{endfdiff_BSFG}
\end{figure}

\begin{figure}[tp]
\includegraphics[angle=270,width=\columnwidth]{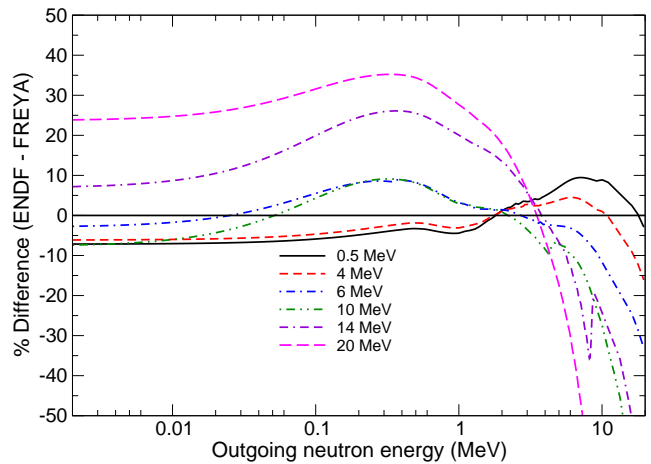}
  \caption[]{(Color online)
The percent difference between the ENDF-B/VII.0 evaluation of the PFNS and our
result with $a = A/e_0$ 
at several representative incident neutron energies.  Results are
shown for 0.5 (solid black), 4 (short dashed red), 6 (dot-dashed blue), 
10 (dot-dot-dashed green), 14 (dot-dash-dashed violet) and 20 (long dashed 
magenta) MeV.}
\label{endfdiff_Aoe0}
\end{figure}

A close inspection of the spectra obtained for 9 and 14~MeV
will reveal abrupt drops in value at the energies
corresponding to the threshold for emission of a second pre-fission neutron,
namely at $\hat{E}_2=E_{\rm n}-S_{\rm n}(^{239}{\rm Pu})$, 3.4 and 
8.4 MeV, respectively.  When the energy of the first pre-fission neutron is 
below  $\hat{E}_2$,
the daughter nucleus is sufficiently excited to make secondary emission
possible.
(These threshold discontinuities are also visible in the spectral differences
shown in Figs.~\ref{endfdiff_BSFG} and \ref{endfdiff_Aoe0}.)
This effect, noted already by Kawano {\em et al.}\ \cite{Kawano01},
grows larger at higher incident energies
where multichance fission is more probable.
Furthermore, when the combined energy of the first two pre-fission neutrons
is below $\hat{E}_3
   =E_{\rm n}-S_{\rm n}(^{239}{\rm Pu})-S_{\rm n}(^{238}{\rm Pu})$ 
the emission of a third pre-fission neutron is energetically possible.
The threshold $\hat{E}_3$ appears as a change in the \code\ slope
relative to the continuous neutron spectrum in ENDF-B/VII.0,
as is visible near 1.35~MeV in the $E_{\rm n}$=14~MeV curve 
in Figs.~\ref{endfdiff_Aoe0} and \ref{endfdiff_BSFG}.
In principle, these onset effects can be measured experimentally
which could thus provide novel quantitative information 
on the degree of multichance fission.

Our final evaluation is based on our fits to $\nubar$ and its energy-energy
covariance, as described above.  The resulting spectral evaluations for both
level density scenarios are
incorporated in different ENDF-type files with their
spectral shapes alone.

To produce the spectral evaluation, requiring fine energy spacings over
the range $10^{-5} \leq E \leq 20$ MeV, from our \code\ results, we fit the 
\code\ PFNS in the regions where either the 
employed bin widths are not sufficiently small (in the region below 0.1 MeV) 
or the statistics are limited (above 6-8~MeV).  
In between, we interpolate the \code\ spectra.  
We also interpolate the spectra around the multichance fission thresholds
where smooth fitting would not be appropriate.

Figures~\ref{endfdiff_BSFG} and \ref{endfdiff_Aoe0} give the difference 
between the present evaluations and the ENDF/B-VII.0 evaluation.  
Both are normalized to unity
at each value of $E_n$.  For incident energies below the threshold
for multichance fission, the difference between the two evaluations
is around 10\% for both methods of calculating the level density.  In the case
where $ a = A/e_0$, our spectra are systematically higher than the 
ENDF/B-VII.0 result.  With the BSFG parameterization of the level density, the
FREYA spectra are systematically lower than the ENDF-B/VII.0 evaluation.   
As the incident energy increases up to $\sim 14$ MeV, in both cases the
\code\ spectra are generally lower than ENDF/B-VII.0 for outgoing energies 
between 0.01 and 3 MeV because pre-fission neutron emission lowers 
the effective temperature of the fragments when fission occurs.  
The \code\ evaluation has
a higher-energy tail than ENDF/B-VII.0 for all $E_n$, 
giving the \code\ results a higher average energy, particularly for the BSFG.

As noted in the discussion of Figs.~\ref{nspectraBSFG} and \ref{nspectra_comp},
contributions from pre-fission neutron emission change the calculated
spectral slope at $E=E_n - S_n$.
Although the Madland-Nix (Los Alamos model) evaluation 
\cite{MadlandNix} includes an average result for multichance fission, 
the spectral shape for pre-fission emission is assumed 
to be the same as that of prompt neutron emission (evaporation) post fission.
Thus the ENDF/B-VII.0 evaluations are always smooth over the entire outgoing
energy regime, regardless of incident energy, while the \code\ evaluations
reflect the changes due to pre-fission emission.
The slope changes at the multichance fission thresholds
in the \code\ spectra are evident for the $E_n = 6$, 10 and 14 MeV difference 
curves at 0.35, 4.35 and 8.35 MeV, respectively.  We note that the threshold
at 0.35~MeV is not well defined for $a = A/e_0$ while it is exaggerated
for the BSFG.

\begin{figure}[tp]
\includegraphics[angle=270,width=\columnwidth]{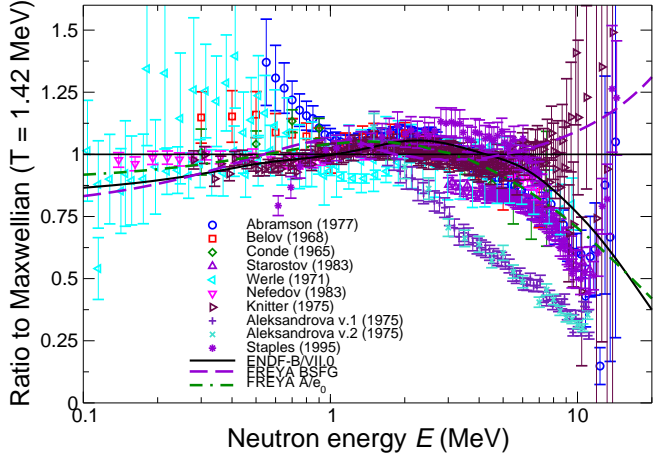}
  \caption[]{(Color online)
The ratios of ENDF-B/VII.0 (solid), FREYA BSFG (long dashed), and 
FREYA $a =A/e_0$
(dot-dashed) relative to a Maxwell distribution with $T = 1.42$~MeV.  The data
from Refs.~\protect\cite{AbramsonPu,BelovPu,CondePu,StarostovPu,WerlePu,NefedovPu,KnitterPu,AleksandrovaPu,StaplesPu} are also shown as ratios to the same Maxwellian.
}
\label{Maxwell}
\end{figure}

Figure~\ref{Maxwell} shows the ratio of our 0.5~MeV results for the $a = A/e_0$
and BSFG scenarios to a Maxwell distribution with $T = 1.42$~MeV.  
The ENDF-B/VII.0 ratio is also shown, along with data from Refs.~\cite{AbramsonPu,BelovPu,CondePu,StarostovPu,WerlePu,NefedovPu,KnitterPu,AleksandrovaPu,StaplesPu}.  As expected from the comparison in Figs.~\ref{endfdiff_BSFG} and 
\ref{endfdiff_Aoe0}, the FREYA ratios bracket the low energy ENDF-B/VII.0 
ratio.  At $E > 0.3$~MeV, the dip relative to ENDF-B/VII.0 around 1~MeV 
followed by a bump at $E \sim 3$~MeV and subsequent decrease in the percent
difference in Fig.~\ref{endfdiff_BSFG} for the BSFG is seen reflected in the 
ratio relative to the Maxwellian.  At higher outgoing energies, the BSFG
result is higher than the Maxwellian.  On the other hand, the $a =A/e_0$ result
is similar to that of ENDF-B/VII.0 but with a slightly lower average energy.
Since there is a great deal of scatter in the data over the entire energy
range, any conclusion about the quality of the fits with respect to the 
spectral data is difficult.

\subsection{Covariances}

We can calculate covariances and correlation coefficients between 
the optimal model parameter values 
as well as between the various output quantities using Eq. (\ref{eq:defofcov}).
The covariance between two parameters $m_k$ and $m_{k'}$ is
\beq\label{sigmakk}
\Sigma_{kk'}\ \equiv\ 
\prec(m_k-\prec m_k\succ)(m_{k'}-\prec m_{k'}\succ)\succ\ .
\eeq
The diagonal elements, $\Sigma_{kk}$ are the variances $\Sigma_{m_k}^2$,
representing the squares of the uncertainty
on the optimal value of the individual model parameter $m_k$,
while the off-diagonal elements give the covariances between
two different model parameters.
It is often more instructive to employ the associated 
{\em correlation coefficients},
$C_{kk'}\equiv\Sigma_{kk'}/[\Sigma_{m_k}\Sigma_{m_{k'}}]$,
which is plus (minus) one for fully (anti)correlated  variables 
and vanishes for entirely independent variables.
The correlation coefficient between the energy independent parameters 
$e_0$ and $x$ is $C_{e_0,x} = 0.985$, a near-perfect correlation, 
for $a = A/e_0$, while  $C_{e_0,x} \approx -0.7$,
a strong anticorrelation, for the BSFG level density.  

\begin{figure}[tp]
\includegraphics[angle=270,width=\columnwidth]{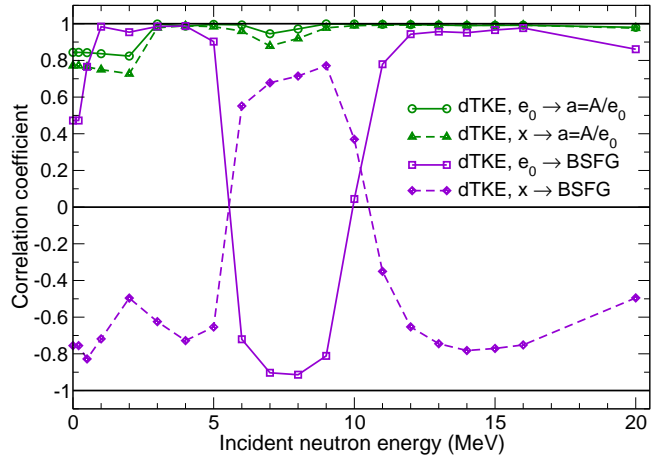}
  \caption[]{(Color online)
The correlation coefficients between $d{\rm TKE}(E_n)$ 
and either $e_0$ (circles) or $x$ (triangles) for $a = A/e_0$
and $e_0$ (squares) or $x$ (diamonds) for the BSFG level density.}
\label{corrtable}
\end{figure}

The correlation coefficients $C_{d{\rm TKE},x}$ and 
$C_{d{\rm TKE},e_0}$ are shown in Fig.~\ref{corrtable}.
When $a = A/e_0$, 
the parameters are all strongly correlated although $C_{d{\rm TKE},e_0}$ is
slightly larger than $C_{d{\rm TKE},x}$.
On the other hand, 
the BSFG inputs are either strongly correlated or anticorrelated with
the sign of the correlation changing near the second and third-chance fission
thresholds.  The anticorrelation of $C_{d{\rm TKE},x}$ and 
$C_{d{\rm TKE},e_0}$ with each other is consistent with the anticorrelation in 
$C_{e_0,x}$ mentioned previously and may
be related to the larger value of $x$ in this case.

\begin{figure}[tp]
\includegraphics[angle=270,width=\columnwidth]{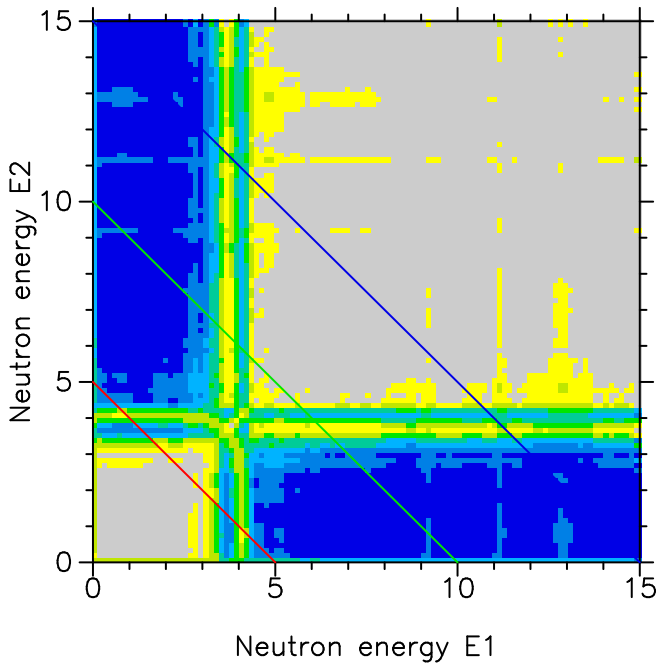}
  \caption[]{(Color online)
Contour plot of the correlation coefficient (see Eq.\ \ref{eq:defofcov})
between the spectral strengths at two different energies, 
$C_{E_1,E_2}$, as obtained for $E_n=0.5$~MeV with the BSFG result.
The correlation changes from values near +1 in the light regions 
(lower-left and central regions) to values near -1 in the dark regions
(near the two axes).  Each of the three straight lines connects points
at which the two neutrons have the same
combined energy, $E_1+E_2=5, 10, 15$~MeV.
}
\label{corr_spec_BSFG}
\end{figure}

\begin{figure}[tp]
\includegraphics[angle=270,width=\columnwidth]{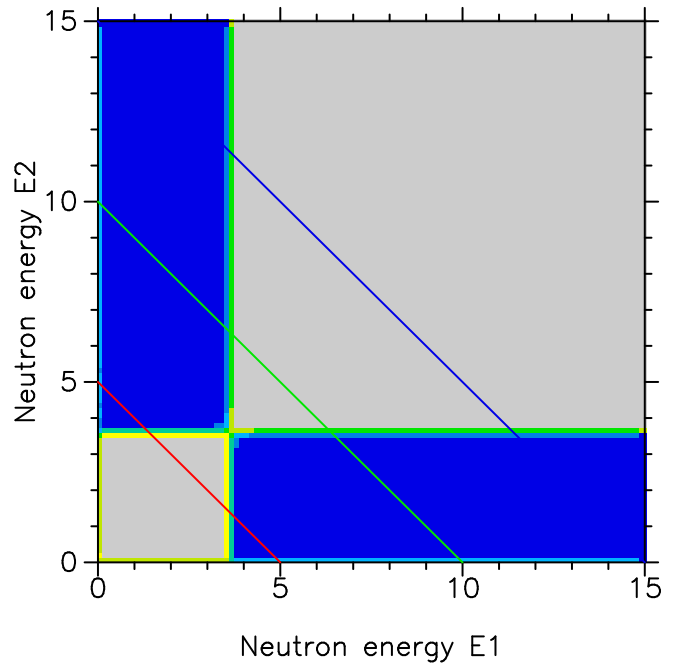}
  \caption[]{(Color online)
Contour plot of the correlation coefficient (see Eq.\ \ref{eq:defofcov})
between the spectral strengths at two different energies, 
$C_{E_1,E_2}$, as obtained for $E_n=0.5$~MeV with $a = A/e_0$.
The correlation changes rapidly from values near +1 in the light regions 
(lower-left and central regions) to values near -1 in the dark regions
(near the two axes).  Each of the three straight lines connects points
at which the two neutrons have the same
combined energy, $E_1+E_2=5, 10, 15$~MeV.
}
\label{corr_spec_oldLD}
\end{figure}

We may also compute the covariance between the spectral strength at
different outgoing energies $E$ using Eq.~(\ref{eq:defofcov}).
The resulting correlation coefficients $C_{E_1,E_2}$ are shown in
Fig.~\ref{corr_spec_BSFG} and \ref{corr_spec_oldLD} for $E_n=0.5$~MeV.  
We see that $C_{E_1,E_2}\approx1$ (light areas)
when the two specified energies lie on the same side of
the crossover region, while $C_{E_1,E_2}\approx-1$ (dark areas)
when they lie on opposite sides.
The crossover region is from 2.5 to 4~MeV indicates that the spectrum tends 
to pivot around $E\approx3.5$~MeV when the parameter space is explored, 
slightly higher $E$ than in Ref.~\cite{VRPY}. 

The BSFG correlation 
in Fig.~\ref{corr_spec_BSFG} is rather noisy.  In contrast, the crossover
for $a = A/e_0$ in Fig.~\ref{corr_spec_oldLD} is very sharp with threshold-like
boundaries between regions of almost perfect correlation and anticorrelation.
In our previous paper \cite{VRPY}, we used Eq.~(\ref{aleveldef}) with 
$U = E^*$, thus the results shown there exhibited slightly weaker correlations
than shown in Fig.~\ref{corr_spec_oldLD}.

\begin{figure}[tp]
\includegraphics[angle=270,width=\columnwidth]{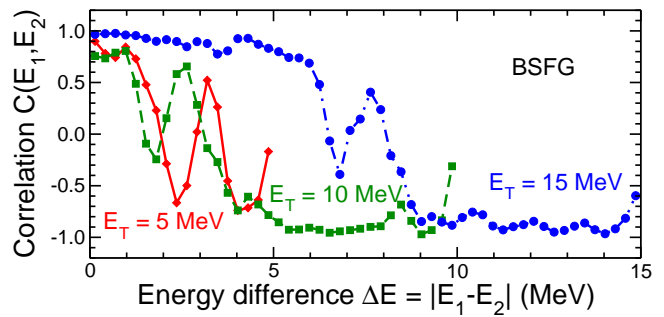}
  \caption[]{(Color online)
The spectral correlation coefficients, $C_{E_1,E_2}$, along 
the three lines of constant combined energy indicated in 
Fig.~\protect\ref{corr_spec_BSFG} for the BSFG result at $E_n = 0.5$ MeV.
}
\label{corr_cut_bsfg}
\end{figure}

\begin{figure}[tp]
\includegraphics[angle=270,width=\columnwidth]{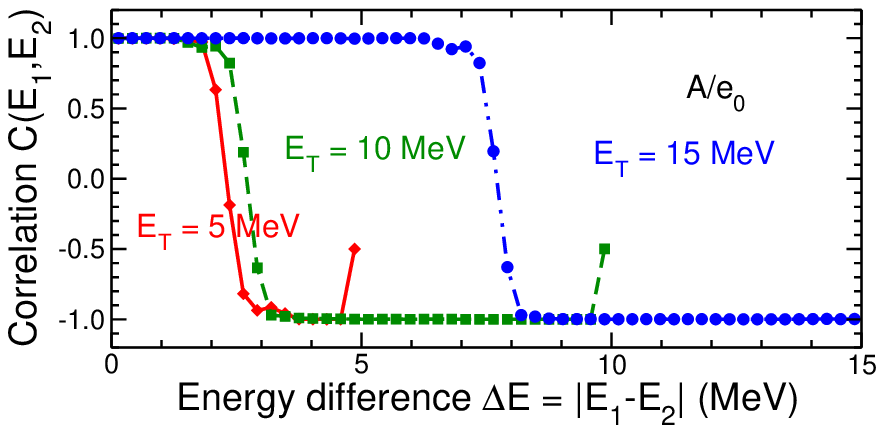}
  \caption[]{(Color online)
The spectral correlation coefficients, $C_{E_1,E_2}$, along 
the three lines of constant combined energy indicated in 
Fig.~\protect\ref{corr_spec_oldLD} for the BSFG result at $E_n = 0.5$ MeV.
}
\label{corr_cut_oldLD}
\end{figure}

Figures~\ref{corr_cut_bsfg} and \ref{corr_cut_oldLD} show 
cuts at constant total neutron energy, $E_k+E_{k'}$. 
Similar results are found for all other incident energies considered.  
The trends are the same for both scenarios, large positive
correlations at low $\Delta E$, becoming negative at larger $\Delta E$.
This is because when model parameters are varied, the spectral shapes pivot 
about a single energy, $E_{\rm pivot} \sim 3.5$~MeV in these calculations.
Thus when both $E_1$ and $E_2$ are less than $E_{\rm pivot}$, the differential
changes are in phase and $C_{E_1,E_2} \sim 1$.  If {\em e.g.} $E_1 < E_{\rm
pivot}$ and $E_2 > E_{\rm pivot}$, differential changes in the spectra give
an anticorrelation.

The fluctuations in $C_{E_1,E_2}$ are very large for the BSFG.  There is a 
shift from negative back to positive and back to negative near the same 
value of $\Delta E$ where $C_{E_1,E_2}$ changes from $+1$ to $-1$ in 
Fig.~\ref{corr_cut_oldLD}.  This coincides with the behavior of the correlation
matrix in Fig.~\ref{corr_spec_BSFG} around the pivot point.  The change in
$C_{E_1,E_2}$ is similar for $E_T = 5$ and 10~MeV, likely because the lines of
constant $E_T$ sit 
at approximately equidistant locations on opposite sides of the pivot point.
In our previous paper, these results were further apart because $E_{\rm pivot}
\sim 2$~MeV \cite{VRPY}.  

There are several possible reasons why the BSFG level density causes 
fluctuations in $C_{E_1,E_2}$.  If $ a = A/e_0$, $ a$ rises 
smoothly with $A$ and is energy independent.  Thus, the temperature is also
independent of incident energy.  Strong correlations are also observed in
other calculations based on average fission models such as Madland-Nix
\cite{ANS}.  Introducing the BSFG parameterization, Eq.~(\ref{aleveldef}), at
fixed $U$, ignoring the pairing energy, as in our previous paper, introduces
fluctuations in $ a$ which soften the linear rise of $ a$ with $A$
and reduce the sharpness of the correlations.  Including the back shift due to
the pairing energy further reduces the correlations, narrowing the peak in the
spectra.  Thus the back shift interferes with the correlations, introducing
the noise shown in Fig.~\ref{corr_spec_BSFG}.

\subsection{Benchmark tests}

There are several standard validation calculations that can be used to test
our PFNS evaluations.  They are critical assemblies which test conditions
under which a fission chain reaction remains stationary, {\em i.e.} exactly
critical; activation ratios which can be used to test the modeling of the flux
in a critical assembly; and pulsed sphere measurements which test the spectra
for incident neutron energies of $\sim 14$~MeV.

To perform these, we replace the $^{239}$Pu
PFNS in the ENDL2011.0 database, identical to that in ENDF-B/VII.0, with our
evaluations obtained with the BSFG level density parameterization and 
$a=A/e_0$.
In this section, we describe these tests and the \code\ results.  As we will
see, the two different treatments of $ a$ do about equally well on these
benchmark tests with no conclusive preference.

\subsection{Validation against critical assemblies}

Critical assemblies, which are designed to determine the conditions
under which a fission chain reaction is stationary,
provide an important quality check on the spectral evaluations.  
The key measure of a critical assembly is 
the neutron multiplication factor $k_{\rm eff}$.
When this quantity is unity, the assembly is exactly critical,
{\em i.e.}\ the net number of neutrons does not change 
so that for every neutron generated, 
another is either absorbed or leaks out of the system.
For a given assembly,
the degree of criticality depends on the multiplicity of prompt neutrons, 
their spectral shape, 
and the energy-dependent cross section for neutron-induced fission.

Plutonium criticality is especially sensitive to the prompt
neutron spectrum because the $^{239}{\rm Pu}(n,f)$ cross section 
rises sharply between $E_{n} = 1.5$ and 2 MeV, near the peak of the 
fission spectrum. 
As a result, increasing the relative number of low-energy neutrons 
tends to decrease criticality, lowering $k_{\rm eff}$,
while increasing the number of higher energy neutrons 
increases criticality. 

\begin{figure}[tp]
\includegraphics[angle=0,width=3.5in]{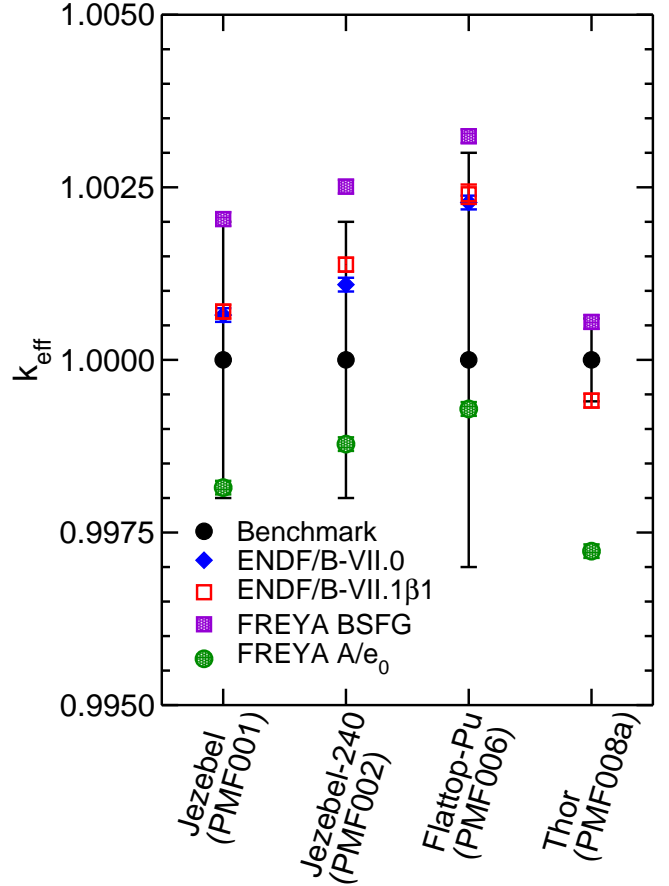}
  \caption[]{(Color online)
Calculated values of $k_{\rm eff}$ for several $^{239}$Pu critical assemblies 
obtained using our fits, 
labeled \code\ (solid squares and circles), in the Mercury code.
The results are compared to calculations using the ENDF-B/VII.0 (blue diamonds)
 and proposed ENDF/B-VII.1 (red squares) databases.}
\label{assembliesfig}
\end{figure}

Figure~\ref{assembliesfig} shows calculations of $k_{\rm eff}$ 
for four different plutonium assemblies from
the criticality safety benchmark handbook \cite{NEA2006}.
Apart from the spectra,
all data used in these calculations were taken from ENDF/B-VII \cite{ENDFb7}. 
We show the $k_{\rm eff}$ for the two level density calculations 
with our evaluated
spectra and $\overline \nu$ from the ENDF-B/VII.0 evaluation.  
The BSFG level density is supercritical $(k_{\rm eff} > 1)$ 
while the $ a = A/e_0$ result is subcritical ($k_{\rm eff} < 1$).

The energy-independent result of Ref.~\cite{VRPY} 
led to values of $k_{\rm eff}$
that were $\approx1.5$ standard deviations away from the measured value for the
Jezebel assembly which is sensitive to fission 
induced by fast ($E_n\approx1$~MeV) neutrons.  
By contrast, our results
for both methods of calculating the level density, 
are within one standard deviation of the measured value and thus
represent some improvement.  

We note that the other inputs to the Jezebel assembly test were highly tuned
to match the $k_{\rm eff}$.  
The fact that replacing only the PFNS without a full reevaluation 
of all the inputs to the criticality tests leads to a result that is
inconsistent with $k_{\rm eff} \equiv 1$ should therefore not be surprising.  
For example, if the average energy of the ENDF-B/VII.0 PFNS is high, the 
$\nubar$ evaluation would be forced higher to counter the effect on $k_{\rm
eff}$.  In addition, the inelastic $(n,n')$ cross section is not well known and
could require compensating changes that affect $k_{\rm eff}$ \cite{Roberto}.

To make further improvements in the evaluations with respect to assemblies, 
it would be useful to have an inline version of \code\
for use in the simulations.

\subsection{Validation against activation ratios}

In the 1970's and 1980's, 
LANL performed a series of experiments using the spectra from the 
 critical assemblies Jezebel (mainly $^{239}$Pu), Godiva, BigTen and Flattop25 
(all primarily enriched uranium)
to activate foils of various materials \cite{activationRatios}.  
The isotopic content of the foils can be radiochemically assessed both
before and after irradiation.  Thus these measurements of the numbers of atoms 
produced per fission neutron are integral test 
of specific reaction cross sections in the foil material.  Conversely,
a well-characterized material can also be used to test the critical assembly 
flux modeling.  
We have simulated several foils ($^{239}$Pu, $^{233}$U, $^{235}$U, $^{238}$U, 
$^{237}$Np, $^{51}$V, $^{55}$Mn, $^{63}$Cu, $^{93}$Nb, $^{107}$Ag, $^{121}$Sb, 
$^{139}$La, $^{193}$Ir and $^{197}$Au) 
which are tests of the $(n,f)$ and $(n,\gamma)$ reactions 
in the Jezebel assembly.
In all cases, 
our simulated values of the activation rates in these foils are either 
consistent with measured values or previous modeling using a modified version
of the ENDF/B-VII.0 nuclear data library \cite{ENDL2009}.  
As the fission cross sections for $^{233}$U, $^{235}$U, $^{238}$U,
 and $^{237}$Np are accurately  known and span all incident neutron energies, 
these tests merely confirm 
our earlier modeling of plutonium critical assemblies.  
The neutron capture cross sections of the other foils 
are important for incident neutron energies less then 1 MeV
but they are not known nearly as well as the fission cross sections.
Thus our Calculated/Experiment ratios scatter around unity in these cases.

Because both the Jezebel and Godiva critical assemblies 
test the fast-neutron spectrum, 
with a significant portion of their neutron fluxes above 5~MeV, 
either might be used to test the high-energy portion of the $^{239}$Pu PFNS.
Unfortunately, none of the tests that have been performed to date are useful 
for testing our \code\ PFNS evaluation.
While many of the studied materials have high ($\geq 10$ MeV) $(n,2n)$
thresholds, the only $(n,2n)$ threshold material tested in Jezebel 
is $^{169}$Tm.  Unfortunately, the $^{169}$Tm $(n,2n)$ cross section is 
poorly known.
There were also experiments with plutonium foils placed in uranium assemblies,
but these tests are not useful for testing the $^{239}$Pu PFNS 
because the foils are too thin for secondary scatterings 
to play a significant role.  
It would particularly interesting to carry out
a new set of $(n,2n)$ foil measurements using well-characterized materials
in a primarily plutonium critical assembly to specifically test
the high-energy portion of the spectrum.


\subsection{Validation with LLNL pulsed spheres}

The ENDL2011.0 database \cite{ENDL2010}, 
including our \code\ evaluation, was tested against 
LLNL pulsed-sphere data, 
a set of fusion-shielding benchmarks \cite{Descalle2008}.  
The pulsed-sphere program,
which ran from the 70's to the early 90's, measured neutron time-of-flight 
(TOF) and gamma spectra resulting from emission of a 14~MeV neutron pulse 
produced by d+t reactions occurring inside spheres composed of a variety of 
materials 
\cite{Wong1972}. Models of the LLNL pulsed-sphere experiments using the 
Mercury Monte Carlo were developed for the materials reported in Goldberg 
{\em et al.} \cite{Goldberg1990,Marchetti1998}.

\begin{figure}[tp]
\includegraphics[angle=270,width=\columnwidth]{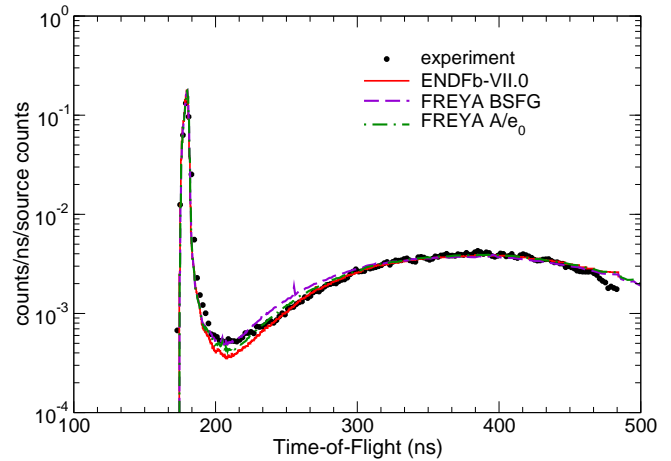}
  \caption[]{(Color online)
The measured LLNL pulsed-sphere test data \protect\cite{Wong1972} (points)
compared to calculations using either ENDF/B-VII.0
or ENDL2011.0 with the present \code\ evaluation included.}
\label{pulsedfig}
\end{figure}

Figure \ref{pulsedfig} compares results of our evaluation 
with the experimental data \cite{Wong1972} 
and calculations based on ENDF/B-VII.0.  The only difference between the Pu
evaluations in these two calculations is the PFNS and associated 
$\overline \nu$, all other quantities remain the same.
Relative to the ENDF/B-VII.0 calculation, in both cases
the \code\ spectra yields significantly better agreement with the data in the 
region around the minimum of the time-of-flight curve at $\approx210$~ns.
However,  it lies somewhat higher than the ENDF/B-VII.0 curve and the data
during the subsequent rise ($240-300$~ns), particularly for the BSFG level 
densities.  Thereafter, the two calculations are practically identical
and agree well with the data until $\approx480$~ns.

Such pulsed-sphere experiments test the PFNS at incident 
energies higher than those probed in critical assembly tests.  
The measured outgoing neutrons
have a characteristic time-of-flight curve, see Fig.~\ref{pulsedfig}. 
The sharp peak at early times is due to 14~MeV neutrons 
going straight through the material without significant interaction.  
The dip at around 200 ns and the rise immediately afterward is caused by 
secondary scattering in the material 
as the neutrons travel out from the center. 
The location and depth of the dip is due to inelastic direct reactions, 
the high-energy tail of the prompt fission neutron spectrum,
and pre-equilibrium neutron emission.  
The last two items are directly addressed by the present evaluation.  
A large part of the secondary interactions are due to neutrons 
that have interacted in the material and are thus less energetic 
than those from the initial 14~MeV pulse. 
At late times, where the results from both evaluations deviate from the data,
the time-of-flight spectrum is dominated by scattering in surrounding
material such as detector components and concrete shielding. 

\subsection{Additional observables}

\begin{figure}[tbp]
\includegraphics[width=\columnwidth]{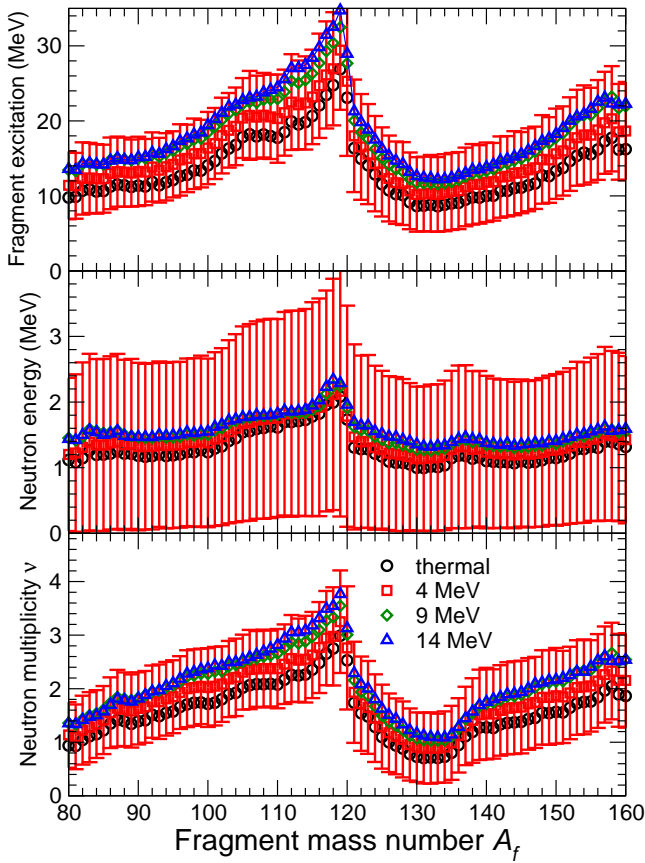}
\caption[]{(Color online)
The total average excitation energy available for neutron emission (top), the 
average neutron kinetic energy (middle) and the average neutron multiplicity
(bottom) as a function of fragment mass number $A_f$ 
for thermal neutrons (black circles), $E_n = 4$ (red squares), 
9 (green diamonds) and 14 (blue triangles) MeV.}
\label{excit-nke-nuvsa-edep}
\end{figure}

\begin{figure}[tbp]
\includegraphics[width=\columnwidth]{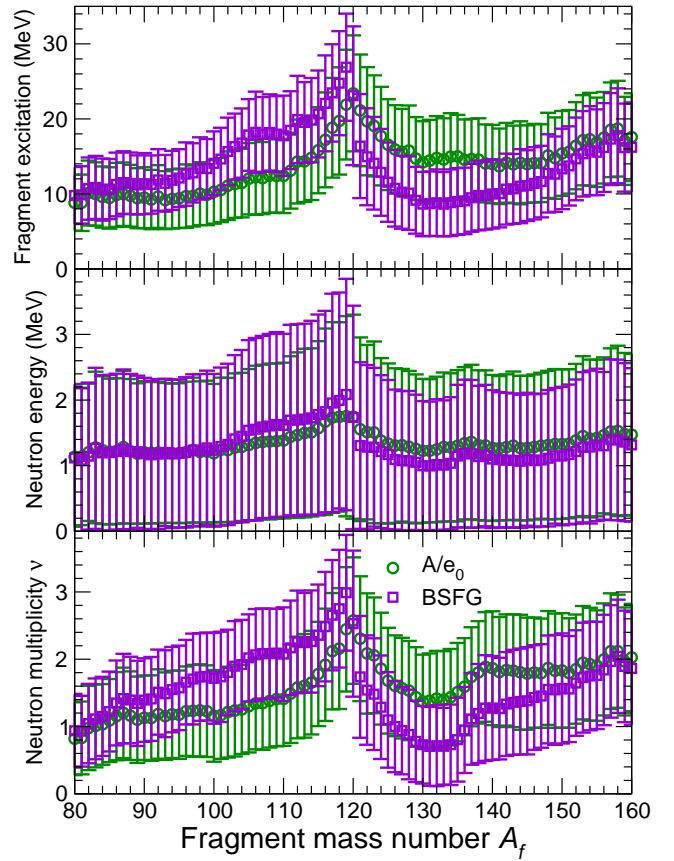}
\caption[]{(Color online)
The total average excitation energy available for neutron emission (top), the 
average neutron kinetic energy (middle) and the average neutron multiplicity
(bottom) as a function of fragment mass number $A_f$ 
for incident neutrons at $E_n = 0.5$ MeV with the level densities calculated
with $ a = A/e_0$ (circles) and the BSFG approach (squares).  The 
variances on the calculated results are shown for $E_n = 4$~MeV only.}
\label{excit-nke-nuvsa-comp}
\end{figure}

The mass-averaged fragment kinetic energies obtained with \code\
are almost independent of the incident neutron energy $E_n$.  
This feature is consistent with measurements made with $^{235}$U and 
$^{238}$U targets over similar ranges of incident neutron energy, 
$0.5\!\leq\!E_n\!\leq\!6$~MeV \cite{HambschNPA491} 
and $1.2\!\leq\!E_n\!\leq\!5.8$~MeV \cite{Vives238}, respectively.  
In both cases, the average TKE
changes less than 1~MeV over the entire energy range.

However, Ref.\ \cite{Vives238} also showed that, 
while the mass-averaged TKE is consistent with near energy independence, 
higher-energy incident neutrons typically give less TKE to
masses close to symmetric fission and somewhat more TKE for $A_H > 140$.
Such detailed information is not available for
neutrons on $^{239}$Pu.  We have therefore chosen to use a constant 
value of $d$TKE at each energy.


\begin{figure}[htbp]
\includegraphics[angle=270,width=\columnwidth]{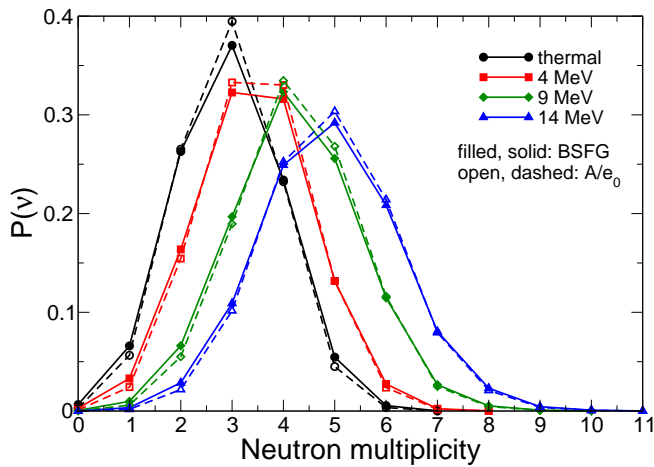}
  \caption[]{(Color online)
The probability for emission of a certain number of neutrons as a function of 
neutron multiplicity for thermal neutrons (black circles), 
$E_n = 4$ (red squares), 9 (green diamonds) and 14 (blue triangles) MeV.  
The filled symbols employ the BSFG level density,
while the open symbols are obtained with $ a = A/e_0$.}
\label{pofnu}
\end{figure}

Neutron observables are perhaps more useful for model validation.
The near independence of TKE($A_H$)
on incident energy implies that the additional energy
brought into the system by a more energetic neutron
will be primarily converted into internal excitation energy.
This is illustrated in the top panel of Fig.~\ref{excit-nke-nuvsa-edep}
which shows the fragment excitation $E^*(A_f)$ for the BSFG level density with
the representative incident energies considered in Fig.~\ref{nspectraBSFG}
(thermal, 4, 9 and 14~MeV): 
$E^*(A_f)$ increases linearly with incident energy.
We note that the form of ${\rm TKE}(A_H)$ (see Fig.~\ref{f:TKEexp})
leads to the familiar sawtooth form of $E^*(A_f)$.

The average kinetic energy of the evaporated neutrons 
is given by $\bar{E}=2T$ for a single emission,
where $T$ is the maximum temperature in the daughter nucleus,
so $T^2\propto E^*-S_n$ for the first emission.
Consequently, $\bar{E}$ should vary relatively little with $A_f$,
as is indeed borne out by the results for $\bar{E}(A_f)$
shown in the middle panel of Fig.~\ref{excit-nke-nuvsa-edep}.
The average outgoing neutron kinetic
energy increases slowly with the incident neutron energy, 
with a total increase of $\approx20$\%  through the energy range shown.  
Although the width of the neutron-energy distribution is given by
$\sigma_E=\bar{E}/\sqrt{2}$ for a single emission,
the resulting width grows faster than that with $E_n$
due to the increased occurrence of multiple emissions and thus 
the appearance of spectral components with different degrees of hardness.

The relatively flat behavior of $\bar{E}(A_f)$ implies
that the neutron multiplicity $\nubar(A_f)$ 
will resemble the fragment excitation energy $E^*(A_f)$,
as is seen to be the case in the bottom panel of
Fig.~\ref{excit-nke-nuvsa-edep}
where the characteristic sawtooth shape of $\nubar(A_f)$ is apparent.  
The number of neutrons from the heavy fragment increases somewhat faster 
with $E_n$ than the number from the light fragment.

Figure~\ref{excit-nke-nuvsa-comp} shows the fragment excitation energy $E^*$,
the kinetic energy of the emitted neutron and the neutron multiplicity, all
as a function of $A_f$ for the BSFG and $ a = A/e_0$ at $E_n = 0.5$ MeV.
The larger $x$
of the BSFG fit gives both a stronger dependence of $E^*$ on $A_f$ and a
sharper `sawtooth' shape, 
more consistent with the data shown in Fig.~\ref{nuvsa_thermal}.
The neutron kinetic energy is not strongly affected by the value of $x$.

\begin{figure}[htbp]	
\includegraphics[width=\columnwidth]{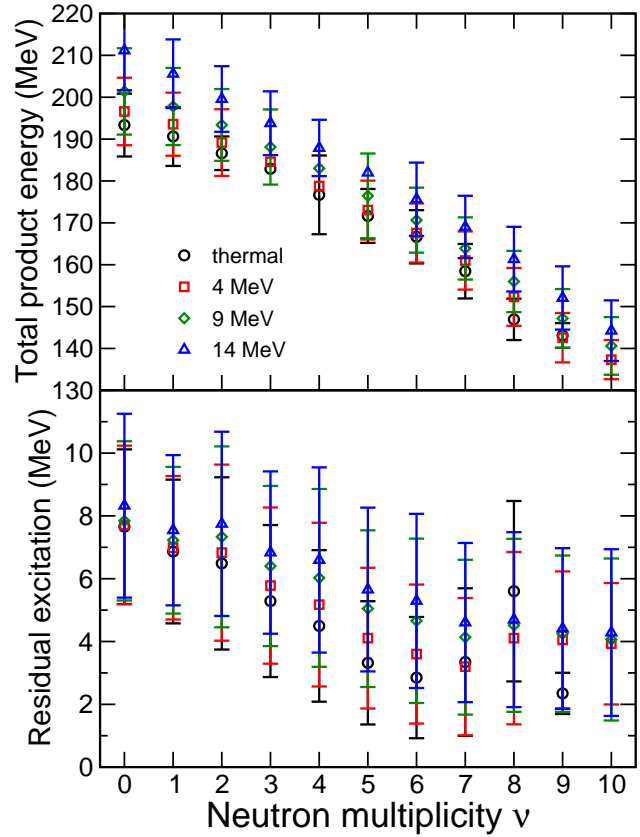}
\caption[]{(Color online)
The combined kinetic energy of the two product nuclei (top) 
and their residual excitation prior to photon emission (bottom)
as functions of the neutron multiplicity $\nu$ 
for thermal neutrons and $E_n = 4, 9, 14$~MeV.
The symbols are at the mean values and the vertical bars 
show the dispersions of the respective distributions.
}\label{prod-res-vsnu}
\end{figure}

\begin{figure}[htbp]	
\includegraphics[width=\columnwidth]{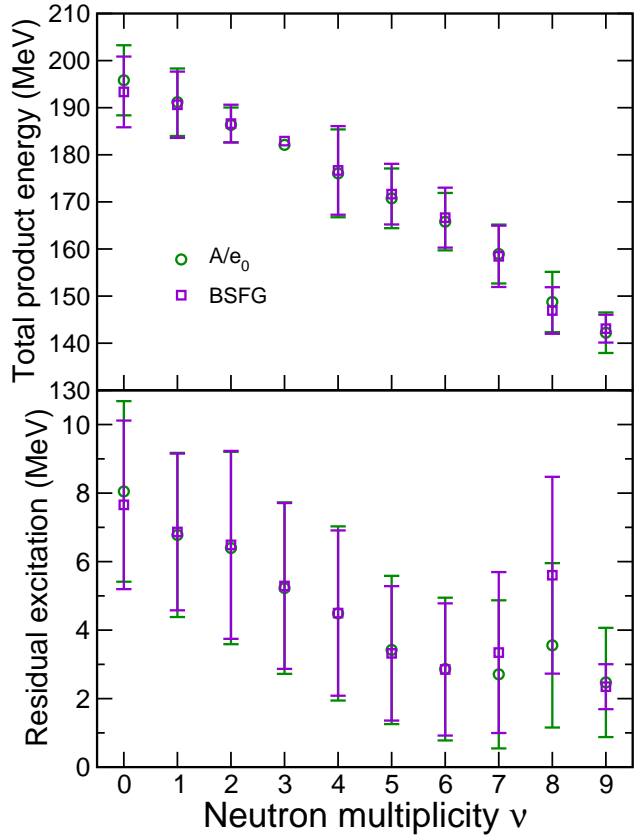}
\caption[]{(Color online)
The combined kinetic energy of the two product nuclei (top) 
and their residual excitation prior to photon emission (bottom)
as functions of the neutron multiplicity $\nu$ 
for $E_n = 0.5$ MeV with the BSFG (squares) and $ a = A/e_0$ (circles).
}\label{prod-res-vsnu-comp}
\end{figure}

New measurements with the fission TPC over a range of incident neutron
energies could provide a wealth of data that could lead to improved modeling.
In addition, calculations of `hot' fission that includes temperature-dependent
shell effects could enhance modeling efforts by predicting trends that could
be input into \code\ and thus test the effects on the PFNS and related 
quantities.

For reasons of computational simplicity, we have chosen to use the same
value of $x$ over the entire energy range considered.  
There are some limited data on thermal neutron-induced fission of $^{235}$U 
\cite{NishioU} and spontaneous fission of $^{252}$Cf \cite{VorobyevCf}
that suggest the light fragment
emits more neutrons than the heavy fragment, 40\% more for $^{235}$U 
\cite{NishioU} and 20\% more for $^{252}$Cf \cite{VorobyevCf}.  
Our BSFG result, $x \sim 1.23$, is consistent with these results.
However, `hotter'
fission could equilibrate the excitation energies of the light and heavy 
fragments which may result in more neutron emission from the heavy fragment,
also reducing the sharpness of the sawtooth pattern. 

Figure~\ref{pofnu} shows the neutron multiplicity distribution $P(\nu)$
for the selected values of $E_n$.
As expected, $\nubar$ increases with $E_n$
and the distribution broadens.
However, each neutron reduces the excitation energy in the residue 
by not only its kinetic energy (recall $\overline E = 2T$)
but also by the separation energy $S_n$ 
(which is generally significantly larger).
Therefore the resulting $P(\nu)$ is narrower than
a Poisson distribution with the same average multiplicity.  These results are
essentially independent of the level density calculation.

The combined kinetic energy of the two resulting (post-evaporation) 
product nuclei is shown as a function of the neutron multiplicity $\nu$
in the top panel of Fig.~\ref{prod-res-vsnu}. 
It decreases with increasing multiplicity,
as one might expect on the grounds that the emission of more neutrons
tends to require more initial excitation energy,
thus leaving less available for fragment kinetic energy.

The bottom panel of Fig.~\ref{prod-res-vsnu} shows the mass dependence of the
average residual excitation energy in those post-evaporation product nuclei.
Because energy is available for the subsequent photon emission,
one may expect that the resulting photon multiplicity
would display a qualitatively similar behavior and thus,
in particular, be anti-correlated with the neutron multiplicity.

There is little sensitivity to the calculated level density in either case, as
shown in Fig.~\ref{prod-res-vsnu-comp} for $E_n = 0.5$ MeV.  This result
shows that the residual energies left over after prompt neutron emission are
not strongly dependent on the temperature.

\section{Conclusion}

We have included both multichance fission and pre-equilibrium emission
into \code\ \cite{VRPY,RV}, 
a Monte-Carlo model that simulates fission on an event-by-event basis.
This has enabled us to perform an extended evaluation of the prompt fission 
neutron spectrum from $^{239}\textrm{Pu}(n,f)$ up to $E_n=20$~MeV.  
Several physics-motivated model parameters have been fitted to the 
ENDF-B/VII.0 evaluation of $\nubar$ and the associated covariance matrix
in two alternate scenarios for the level-density parameterization.

Our testing of these two alternate evaluations was inconclusive: neither 
evaluation performed as well as the ENDF/B-VII.0 evaluation in critical 
assembly benchmarks and, in fact, our two evaluations bracket the 
ENDF-B/VII.0
evaluation.  However, we found improved agreement with the LLNL pulsed sphere 
tests, especially just below the 14 MeV peak in the neutron leakage spectrum, 
for both variants of our evaluation.  Although these mixed results may limit 
the utility of our evaluation in applications, they do give us hope that 
further improvements to the evaluation will either tighten up agreement with 
the critical assemblies or point to other deficiencies in the ENDF-B/VII.0 
$^{239}$Pu evaluation.   

Further investigations will require fitting to other data 
less sensitive to the $\bar{\nu}$ data employed in this work including the 
albeit low quality PFNS data and $\bar{\nu}(A)$ data.

\section*{Acknowledgments}

We wish to acknowledge many helpful discussions with R.~Capote Noy,
M.~Chadwick, T.~Kawano, P.~M{\"o}ller, 
J.~Pruet, W.J.~Swiatecki, P.~Talou, and W.~Younes.
This work was performed under the auspices of the 
U.S. Department of Energy by Lawrence Livermore National Laboratory under 
Contract DE-AC52-07NA27344 (RV, DB, MAD, WEO), by Lawrence
Berkeley National Laboratory under Contract DE-AC02-05CH11231 (JR) 
and was also supported in part by the National 
Science Foundation Grant NSF PHY-0555660 (RV).

\begin{widetext}

\end{widetext}

\end{document}